\documentclass[referee]{jfm}
\usepackage{natbib}
\usepackage{graphicx}
\usepackage{upmath}
\usepackage{amssymb}
\usepackage{amsmath}
\usepackage{amstext}

\newcommand{\Wmax}{\textwidth}

\title[Microstructure and velocity fluctuations]{Microstructure and velocity fluctuations
in sheared suspensions}
\author[G. Drazer$^1$, J. Koplik$^1$, B. Khusid$^2$ and A. Acrivos$^1$]
{G\ls E\ls R\ls M\ls A\ls N\ns D\ls R\ls A\ls Z\ls E\ls R$^1$,\ns J\ls O\ls E\ls L\ns K\ls O\ls 
P\ls L\ls I\ls K$^1$, \ns B\ls O\ls R\ls I\ls S\ns K\ls H\ls U\ls S\ls I\ls D$^2$\ns 
\and A\ls N\ls D\ls R\ls E\ls A\ls S\ns A\ls C\ls R\ls I\ls V\ls O\ls S$^1$}
\affiliation{$^1$The Levich Institute, T-1M, The City College of the
City University of New York,\\ New York, NY 10031, USA \\[\affilskip]
$^2$Department of Mechanical Engineering, New Jersey Institute of Technology,
\\ University Heights, Newark, New Jersey 07102}
\pubyear{2003}
\volume{}
\pagerange{}
\date{\today}
\begin{document}

\maketitle

\begin{abstract}
The velocity fluctuations present in macroscopically homogeneous suspensions of neutrally buoyant, 
non-Brownian spheres undergoing simple shear flow, and their dependence on the microstructure 
developed by the suspensions, are investigated in the limit of vanishingly small Reynolds
numbers using Stokesian dynamics simulations. 
We show that, in the dilute limit, the standard deviation of the velocity fluctuations 
(the so-called suspension temperature) is proportional to the volume fraction, in both 
the transverse and the flow directions, and that a theoretical prediction, which considers only for 
the hydrodynamic interactions between isolated pairs of spheres, is in good agreement with
the numerical results at low concentrations. 
Furthermore, we show that the whole velocity autocorrelation function can be predicted, in the dilute
limit, based purely in two-particle encounters.
We also simulate the velocity fluctuations that would
result from a random hard-sphere distribution of spheres in simple shear flow, and thereby investigate
the effects of the microstructure on the velocity fluctuations. Analogous results are discussed for
the fluctuations in the angular velocity of the suspended spheres. In addition, we present the
probability density functions for all the linear and angular velocity components, and for three different
concentrations, showing a transition from a Gaussian to an Exponential and finally to a Stretched 
Exponential functional form as the volume fraction is decreased. 

The simulations include a non-hydrodynamic repulsive force between the spheres which, although extremely
short ranged, leads to the development of fore-aft asymmetric distributions for large enough volume
fractions, if the range of that force is kept unchanged. On the other hand, we show that, although
the pair distribution function recovers its fore-aft symmetry in dilute suspensions, it remains anisotropic
and that this anisotropy can be accurately described by assuming the complete absence of any permanent
doublets of spheres.   

We also present a simple correction to the analysis of laser-Doppler velocimetry measurements,
that only takes into account the mean angular rotation of the spheres in the vorticity direction, 
and which substantially improves the interpretation of these measurements at low volume fractions.
\end{abstract}

\section{Introduction}
\label{intro}
The problem of determining the velocity fluctuations in suspensions of non-Brownian solid spheres in 
Stokes flows is one of long-standing difficulty due to the underlying long-range many-body hydrodynamic 
interactions between the suspended particles. 
Even an apparently very simple case, that of determining the dependence on the shear rate of the velocity 
fluctuations in simple shear flows, remains a matter of some controversy \cite[][]{ShapleyAB02}. 
What is clear is that, although the suspension might be homogeneous at macroscopic scales, the continuous 
rearrangements in the suspension microstructure and the corresponding hydrodynamic interactions between 
particles lead to fluctuations in the particle velocities about their mean values, in both the transverse 
and the flow directions.  

In our previous work \cite[][to be referred hereafter as paper I]{acrivos1}, we showed that the dynamics 
of sheared suspensions is chaotic and offered evidence that the chaotic motion is responsible for the loss 
of memory in the evolution of the system. 
This loss of memory, coupled with the fluctuations in the velocity of the spheres, ultimately leads to the 
phenomenon of shear-induced particle diffusion. 

The variance, or the standard deviation (STD), of the velocity fluctuations is the simplest measure of the 
magnitude of such fluctuations and is sometimes referred to as the suspension {\it temperature}, which in 
the case of an anisotropic motion of the suspended spheres would actually be a tensor (covariance matrix).  
The suspension temperature is relevant to the migration and diffusion of particles in shear flows, phenomena 
that occur in a wide variety of natural as well as engineering problems, ranging from the dispersion and 
migration of red blood cells \cite[][]{BishopPIJ02} to the    food industry \cite[][]{CullenDOO00,gotz2003}, hence 
it is important to determine its properties.  
In particular, we are interested in the dependence of the velocity fluctuations on the concentration and 
microstructure of the suspension.  
Unfortunately, and in contrast to the well-studied sedimentation problem, velocity fluctuations in sheared 
suspensions have received little attention thus far.
In recent experiments, \cite{AverbakhSNS97,ShaulyANS97,LyonL98i} and \cite{ShapleyAB02} used laser-Doppler 
velocimetry (LDV) to measure the velocity fluctuations in concentrated suspensions of monodisperse spheres. 
\cite{AverbakhSNS97} and \cite{ShaulyANS97} measured such velocity fluctuations in rectangular ducts and 
found that the STD's, both along and transverse to the flow, depend linearly on the shear rate (or on the 
maximum velocity inside the rectangular channel), as expected in the Stokes limit
\footnote{Let us note that in these experiments, the main contribution to the measured STD's were not actual 
fluctuations but migration and angular velocities of the suspended particles, as noted by the authors.}. 
\cite{LyonL98i} measured the time-averaged local STD in the direction of the flow, for concentrated suspensions 
flowing in a two-dimensional rectangular channel, and also found a linear dependence on the volumetric flow rate. 
\cite{ShapleyAB02} presented the first detailed measurements of the velocity fluctuations in both the transverse 
and the flow directions, as well as of the dependence of the suspension temperature on the volume fraction and 
shear rate, for suspensions undergoing simple shear flow in a Couette device.  
Their results stress the difficulties encountered in such measurements and the discrepancy among different 
experimental results.  They found a highly anisotropic temperature tensor, with the magnitude of the fluctuations 
in both transverse components of the velocity smaller than that in the direction of the bulk flow. 
\cite{ShapleyAB02} also found that the temperature is not monotonically increasing with volume fraction, as is 
usually expected, but shows a different behavior for each of its components.  
Specifically, the component of the temperature in the direction of flow was found to decrease with concentration, 
that in the direction of the gradient stayed constant, while that in the vorticity direction initially increased 
in magnitude with increasing concentrations and then decreased for concentrations larger than $40\%$.  
Finally, and most surprisingly, \cite{ShapleyAB02} found that, whereas the fluctuations in the 
direction of the flow increased linearly with shear rate (as expected for any flow in the Stokes regime), the STD in 
the vorticity direction increased non-linearly, while that in the gradient direction slightly decreased with shear rate.

In terms of the suspension spatial structure, although a larger body of experimental
information exists concerning the microscopic structure developed by suspensions of monodisperse, 
non-Brownian spheres undergoing linear shear flow, no measurements of how the velocity 
fluctuations are affected by the suspension microstructure appear to have been conducted 
thus far. Recall that the experimental work of \cite{Gadala-MariaA80} provided, for the first time, clear 
evidence that concentrated suspensions of monodisperse, non-Brownian spheres develop 
an anisotropic structure when sheared. They showed that, when the direction of shear
was reversed, the shear stress measured in a parallel plate device underwent a transient
response not present when the shearing was started again in the same direction, and thereby concluded
that the underlying structure was not only anisotropic but asymmetric
under reversal of the flow direction, i.e. fore-aft asymmetric. 
Their oscillatory experiments showed similar results, in that the measured dynamic 
viscosity $\mu^\prime$, although independent of the frequency of oscillation at low frequency, was consistently smaller 
that the shear viscosity $\mu$ of the suspension, stressing again the presence of a microscopic structure
induced by the shear. 
In recent experiments, \cite{KolliPG02} used a parallel ring geometry that allowed them to measure 
the normal stress response to shear reversal in concentrated suspensions, in addition to measuring 
the shear stress behavior, and found a transient response in both the normal and the shear stresses 
when the shear was restarted in the opposite direction.
Moreover, the absolute value of both the normal and the shear stresses changed at the very instant of flow reversal, 
which means that the fore-aft asymmetry in the microstructure alone is not enough to explain 
the observed response in the stress upon shear reversal, 
but that non-hydrodynamic forces must also have been acting on the system, either in the
form of repulsion forces or of rough contacts between spheres.
Even more complicated shear stress responses, including shear-induced ordering, has been recently reported 
at large concentrations ($\phi > 50\%$), probably corresponding to a regime in which non-hydrodynamic 
interactions dominate the behavior of the system \cite[][]{VoltzNHR02}.
The first direct observations of the microscopic structure developed by dilute suspensions 
($\phi = 1\% - 5\%$) undergoing shear were presented by \cite{HusbandG87}, who measured 
in a Couette device the relative distribution of spheres centers in the plane of shear, 
and then by averaging over many realizations, found an anisotropic 
but fore-aft symmetric distribution of close particles. 
The anisotropy was attributed to the presence of pairs of spheres rotating around each other
forming {\it permanent doublets}. 
On the other hand, in similar experiments, \cite{ParsiG87} showed
that concentrated suspensions ($\phi = 40\% - 50\%$) do exhibit fore-aft asymmetry,
with a larger probability of finding pairs of spheres oriented on the approaching side
of the reference particle, and attributed this asymmetric distribution to either 
the intrinsic roughness of the spheres or to the presence of a non-hydrodynamic 
repulsive force between particles. 
More recently, \cite{RampallSL97} used a substantially improved flow-visualization 
technique to measure the pair distribution function of dilute suspensions ($\phi = 5\% - 15\%$)
undergoing simple shear flow in a shear tank apparatus, and showed that, contrary to
the results of \cite{HusbandG87}, there is a depletion of {\it permanent doublets} moving in the 
region of closed streamlines, and that even for concentrations as small as $5\%$ the distribution 
is fore-aft asymmetric. Using the surface roughness model of \cite{daCunhaH96}, and assuming that no 
particles formed permanent doublets, \cite{RampallSL97} were able to reproduce the qualitative trends in the pair
distribution function, but the predicted depletion of spheres in the regions aligned with 
the flow was much larger than that observed. 

In view of the contradictory results outlined before, it is clear that numerical simulations,
specifically Stokesian dynamics which are well suited for studying low-Reynolds number flows
of suspensions \cite[][]{Brady01}, offer an important complement to experiments, in that they 
can provide detailed, microscopic information that is not accessible via currently available 
experimental techniques. In their original work describing their Stokesian dynamics method, 
\cite{BossisB84} showed that the pair distribution function of unbounded suspensions undergoing 
simple shear flow had an angular dependence, with the microstructure being no longer fore-aft 
symmetric, and that very few particles were oriented in the receding side of the reference sphere. 
In paper I, we also showed such a break in the fore-aft symmetry in the presence of large 
non-hydrodynamic forces acting between the spheres but, for sufficiently
small repulsion forces, we found that, although anisotropic, 
the pair distribution function becomes fore-aft symmetric, as expected for purely
hydrodynamic interactions. To our knowledge, however, as yet no systematic numerical investigation
has been made of the velocity fluctuations in sheared suspensions and their dependence on 
the underlying microstructure of the suspension.
 
It is the purpose of this paper to investigate the velocity fluctuations present in a macroscopically
homogeneous, unbounded suspension of neutrally buoyant, non-Brownian sphe\-res subject to a simple shear 
flow in the limit of vanishingly small Reynolds numbers using Stokesian dynamics, and their dependence 
on the microstructure developed by the suspensions. 
First, we shall focus on the anisotropic, but fore-aft 
symmetric, distribution of close pairs observed in dilute suspensions, and show how it can be accurately 
described assuming the absence of permanent doublets of spheres, as first suggested by \cite{RampallSL97}. 
We shall also point out that, although the use of a non-hydrodynamic interparticle force 
of extremely short range will yield symmetric distributions, as was reported in paper I, 
the suspensions develop a fore-aft asymmetry for large enough volume fractions if the range
of that force is kept unchanged. 
Then, we shall show that the pair distribution function $g_{\text{\tiny BG}}(r)$ obtained by 
\cite{batchelor72b} in the dilute limit, accurately describes the microstructure in sheared suspensions, 
in particular the divergence of the probability density of finding pairs of spheres nearly touching one 
another, even though it does not account for the observed depletion of closed pairs. 
Then, by making use of the pair distribution function $g_{\text{\tiny BG}}(r)$, 
we shall compute all the temperature components in the dilute limit by 
numerically integrating the expressions given by \cite{batchelor72a} for the particle velocities 
of two freely suspended spheres interacting only through hydrodynamic forces in the presence of
a simple shear flow, and then compare the results with those obtained from the numerical simulations. 
Some general properties of the temperature tensor valid for isotropic pair distribution functions will also 
be discussed. The velocity fluctuations at larger concentrations show the effect of the anisotropic structure
developed by the flow in that some symmetries of the temperature tensor are lost. We also simulate the 
velocity fluctuations that would result from a random hard-sphere spatial distribution of particles in a simple 
shear flow, and thereby are able to further investigate the effects of the microstructure, both its angular 
and radial dependence, on the temperature tensor. 
In addition, the numerical simulations provide a full picture 
of the velocity fluctuations and to this end we shall present the probability density functions for all
the linear and angular velocity components at three different concentrations, showing a transition from a
Gaussian to an Exponential and finally to a Stretched Exponential form as the volume fraction is decreased.
Finally, we shall propose a simple correction to the data reduction analysis of the velocity measurements 
in LDV experiments, that only depends on the mean angular rotation of the spheres in the vorticity direction, 
and which substantially improves the interpretation of the LDV measurements at low volume fractions. 

\section{Simulation method: Stokesian dynamics}
\label{simulation}
We investigate the behavior of suspensions of non-Brownian particles subject to simple shear using the 
method of Stokesian dynamics. 
A detailed description of the method is given in a review by \cite{brady88}, and the specifics of our 
simulations were already discussed in paper I, hence only a brief discussion is presented here.  
The method accounts for the hydrodynamic forces between solid spheres undergoing simple shear, 
characterized by a shear rate $\dot\gamma$, in the limit of zero Reynolds number. 
In order to simulate the behavior of infinite suspensions, periodic boundary conditions in all directions 
are imposed.  
The simulated cubic cell contains a fixed number of spheres $N$, related to the volume fraction 
$\phi$ by $\phi = (4\pi a^3/3) N/V$, where $V$ is the volume of the cell.  
Interactions between particles more than a cell apart are included using the Ewald method. 
A typical simulation consisted of $N=64$ particles sheared over a period of time $t \sim 100 \dot\gamma^{-1}$, 
and all measurements to be reported in this work are for strains $\dot\gamma t$ in excess of $50$, when the 
system has reached its steady or fully developed state.  
The motion of the particles was integrated using a constant time step $\Delta t = 10^{-3} \dot\gamma^{-1}$.  
The results are averaged over $N_c \sim 100$ different initial configurations, with each initial configuration 
corresponding to a random distribution of non-overlapping spheres in the simulation cell, using the random-phase 
average method proposed by \cite{marchioro2001}. 
In what follows, we shall express all the variables in dimensionless units, using the radius 
of the spheres $a$ as the characteristic length and $\dot\gamma^{-1}$ as the characteristic time.

In a suspension of monodisperse spheres undergoing simple shear the separation between 
spheres may become exceedingly small during two-particle collisions (less than $10^{-4}$ of 
their radius), and the effects of surface roughness or small Brownian displacements 
cannot be neglected. Usually, a short-ranged, repulsive force is introduced between 
the spheres to qualitatively model the effect of these non-hydrodynamic interactions,
with the numerical advantage of preventing any overlaps during close encounters between particles. 
As in paper I, we used the following standard expression for the repulsive interparticle force, 
\begin{equation}
\label{inteparticle}
{\bf F}_{\alpha \beta}= \frac{F_0}{r_c}
\frac{{\rm e}^{-\epsilon/r_c}}{1- {\rm e}^{-\epsilon/r_c}} 
{\bf e}_{\alpha \beta} \, ,
\end{equation}
where 6$\pi \mu a^2 \dot\gamma{\bf F}_{\alpha \beta}$,
with $\mu$ being the viscosity of the suspending liquid,
is the force  exerted on sphere $\alpha$ by sphere $\beta$,
$F_0$ is a dimensionless coefficient reflecting the magnitude of this force,
$r_c$ is the characteristic range of the force,
$\epsilon$ is the distance of closest approach between the surfaces
 of the two spheres divided by $a$,
and ${\bf e}_{\alpha \beta}$ is the unit vector connecting their centers
pointing from $\beta$ to $\alpha$. 

The effect of the characteristic range of the interparticle force $r_c$ on the
microscopic structure of the suspension was discussed in paper I. First, we showed
that the minimum separation reached by colliding spheres, and therefore the first 
peak in the pair distribution function, is strongly affected by the range of the 
interparticle force in that, as $r_c$ increases, the minimum separation between 
neighboring particles also increases. Then, we showed that, in general, the presence of a repulsive 
force breaks the fore-aft symmetry of the particle trajectories 
in a simple shear flow. However, we also showed that the symmetry is recovered, 
for small enough values of the force range, $r_c\sim10^{-4}$, at least in the sense that
no asymmetry was observed in the angular dependence of the numerically computed pair distribution function.   
In this work, we use this small range for the interparticle force, $r_c\sim10^{-4}$.

\section{Microscopic structure induced by the shear}
\label{structure}

The investigation of the microscopic structure developed by suspensions undergoing shear flow followed 
the pioneering work by \cite{batchelor72b}, where an expression was derived for the pair distribution 
function $g({\boldsymbol r})$ in the dilute limit. 
This function is related to $P({\boldsymbol r}|{\boldsymbol r_0})$, the probability of finding a sphere 
with its center at $\boldsymbol r$ given that there is a sphere with its center at ${\boldsymbol r_0} = 0$,  
by $P({\boldsymbol r}|{\boldsymbol r_0}) = \phi g({\boldsymbol r}) (3/4\pi)$. 
Recall that, even a random hard-sphere distribution leads to correlations in the position of any two 
particles due to excluded volume effects, and typically displays a liquid-like microstructure at high 
volume fractions. 
But, in addition to these excluded volume effects, hydrodynamic interactions between spheres lead to 
surprising results. 
Specifically, \cite{batchelor72b} showed that the pair distribution function is an isotropic function 
of the distance between the two spheres, i. e. $g({\boldsymbol r}) = g(r)$, and that it diverges as 
$r \to 2$, which means that pairs of particles are substantially more likely to be found near 
contact in a sheared suspension than in a random hard-sphere distribution. 
In turn, this implies a high correlation in the position of the spheres that is not present in a random 
hard-sphere configuration.
The expression for $g({\boldsymbol r})$ derived by \cite{batchelor72b} in the dilute approximation is,
\begin{equation}
\label{grbatchelor}
g_{\text{\tiny BG}}(r) = \frac{1}{1-A(r)} \exp \left\{ \int^\infty_r \frac{3}{\xi} \frac{B(\xi)-A(\xi)}{1-A(\xi)} d\xi \right\},
\end{equation}
where the mobility functions $A$ and $B$ are functions only of $r$. 
Here, we shall use the expressions for these functions given by \cite{daCunhaH96}, who divided the interval $r>2$ 
into three different regions (see \cite{daCunhaH96} for details on how they obtained the expressions for 
$A$ and $B$ in each region). 
Specifically: a) within the {\it lubrication} region $2 < r \leq 2.01$, 
\begin{eqnarray}
\label{ABlubrication}
\nonumber
A &=& \frac{(16.3096-7.1548 r)}{r},
\\ \nonumber
B &=& \frac{2}{r} \frac{(0.4056 L^2 + 1.49681 L - 1.9108)}{(L^2 + 6.04250 L + 6.32549)},
\end{eqnarray}
where $L=-\ln(r-2)$, 
\newline
b) within the {\it intermediate} region $2.01 < r \leq 2.5$,
\begin{eqnarray}
\label{ABintermediate} 
\nonumber
A &=& -4.3833+17.7176 \frac{1}{r}+ 14.8204 \frac{1}{r^2}- 92.4471 \frac{1}{r^3} - 46.3151 \frac{1}{r^4}+ 232.2304 \frac{1}{r^5}, 
\\ \nonumber
B &=& -3.1918+12.3641 \frac{1}{r}+ 11.4615 \frac{1}{r^2}- 65.2926 \frac{1}{r^3} - 36.4909 \frac{1}{r^4}+ 154.8074 \frac{1}{r^5},
\end{eqnarray}
and c) within the {\it far-field} region $r > 2.5$, 
\begin{eqnarray}
\label{ABfarfield}
\nonumber
A &=& \frac{5}{r^3} - \frac{8}{r^5} + \frac{25}{r^6} - \frac{35}{r^8} + \frac{125}{r^9} 
- \frac{102}{r^{10}} + \frac{12.5}{r^{11}} +\frac{430}{r^{12}}, 
\\ \nonumber 
B &=& \frac{1}{3} \left(  \frac{16}{r^5}+\frac{10}{r^8}-\frac{36}{r^{10}}-\frac{25}{r^{11}}-\frac{36}{r^{12}}\right). 
\end{eqnarray}

To be precise, these results for the pair distribution function $g_{\text{\tiny BG}}(r)$, 
apply only to particles that are initially far from each other, and therefore spatially uncorrelated. 
But, in the limit of purely 
hydrodynamic interactions and very dilute suspensions (no three-particle interactions), 
there is a region of close trajectories in $\boldsymbol r$-space, where pairs of particles remain 
correlated at all times forming permanent doublets \cite[][]{batchelor72a}. 
In this case, accounting for the probability distribution of particles forming permanent doublets would require the 
knowledge of the initial distribution of the particles. On the other hand, 
the presence of any non-hydrodynamic interaction, 
such as roughness or repulsion forces, or the existence of three-particle collisions, would generate a transfer of 
particles across the streamlines and therefore remove the need for specifying the initial distribution of spheres. 
Even so, to obtain the probability distribution 
would still require the full knowledge of the transfer process (the combination of three-particle collisions and 
non-hydrodynamic interactions), and the solution of the corresponding boundary value problem.
\begin{figure}
\centering
\includegraphics*[width=\Wmax]{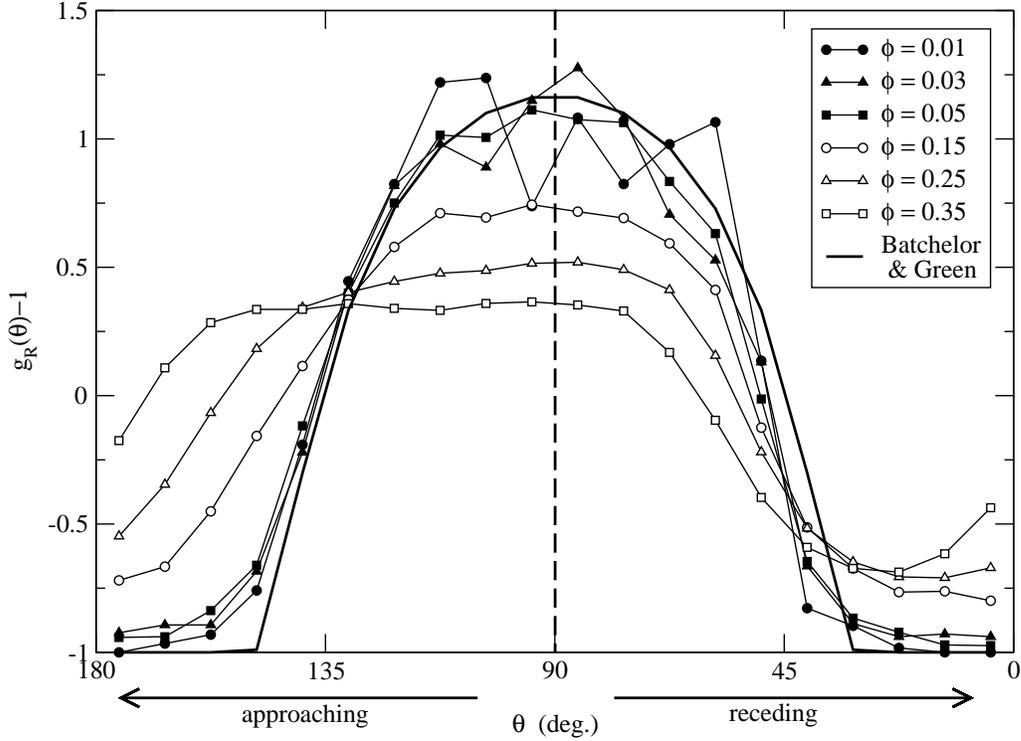}
\caption[]{Normalized angular distribution function $g_R(\theta)$ for pairs of particles. 
The distance between the pairs lies in the range $2 < r < 2.01$ ($R=2.01$). 
Different curves are for different volume fractions. 
All simulations were performed with $N=64$, $N_c=100$, $F_0=1$ and $r_c=10^{-4}$. 
The solid line is obtained using the pair distribution function given in Eq.\ref{grbatchelor} 
for the region outside the closed streamlines and assuming zero probability of finding a pair 
forming a permanent doublet (see text for a more detailed explanation).}
\label{gtheta}
\end{figure}

This implies that the pair distribution function may actually be anisotropic simply due to the 
distribution of pairs forming permanent doublets in that, although the distribution of particles 
outside the region of closed streamlines does not have an angular structure in the dilute limit, 
as shown by \cite{batchelor72b}, the distribution of particles in the region of closed streamlines, 
which extends to $r\to \infty$, may actually render the complete pair distribution function 
anisotropic. 
In fact, in paper I we showed that, although for exceedingly short ranged repulsion forces, 
$r_c\sim10^{-4}$, the pair distribution function for close particles recovered its expected 
fore-aft symmetry, it remained anisotropic. 
In figure \ref{gtheta} we present $g_R({\theta})$, the angular dependence of the pair distribution 
function of pairs closer than a certain distance $R$, as defined in paper I, for different 
concentrations of the suspension (as mentioned in section \S\ref{simulation}, all numerical results
are for a range $r_c=10^{-4}$ of the interparticle force, which is the smallest value of the 
force range simulated in paper I). 
\begin{figure}
\centering
\includegraphics*[width=\Wmax]{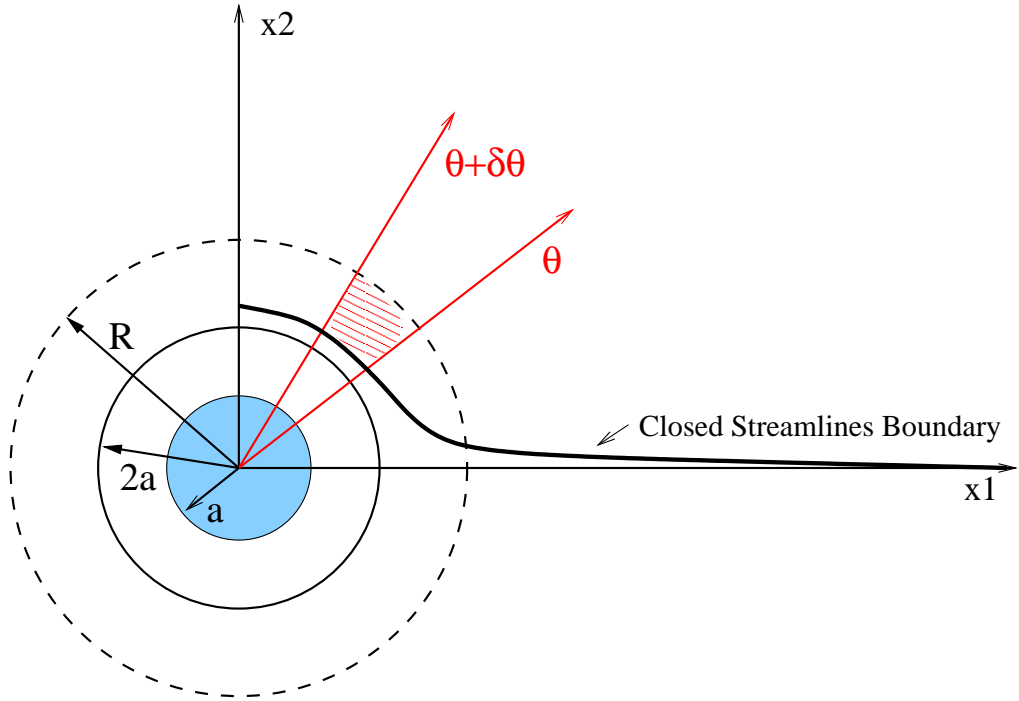}
\caption[]{Schematic representation of the computation of $g_R(\theta)$ using the regions of open and 
closed streamlines for a pair of interacting spheres \cite[][]{batchelor72a}. 
The solid line represents the intersection of the plane of shear $x_1-x_2$ with the surface 
bounding the region of closed trajectories, which can be formed by rotating this curve about the $x_2$ axis 
together with its mirror image obtained by reflection about the $x_1$ axis. 
The gray circle of radius $a$ represents the reference sphere and the circle of radius
$2a$ encloses the excluded volume. Then, a particle with its center located inside the sphere of 
radius $R$ forms, with the reference sphere, a pair that is closer than $R$, and is included in $g_R(\theta)$. 
For a given angle $\theta$, the distribution of pairs closer than $R$ is calculated by integrating the pair 
distribution function given in Eq. \ref{grbatchelor} only in the shaded 
region, which corresponds to open trajectories only, because the region of closed
trajectories is considered to have a negligible effect due to the depletion of permanent doublets.}
\label{closed}
\end{figure}
\begin{figure}
\centering
\includegraphics*[width=\Wmax]{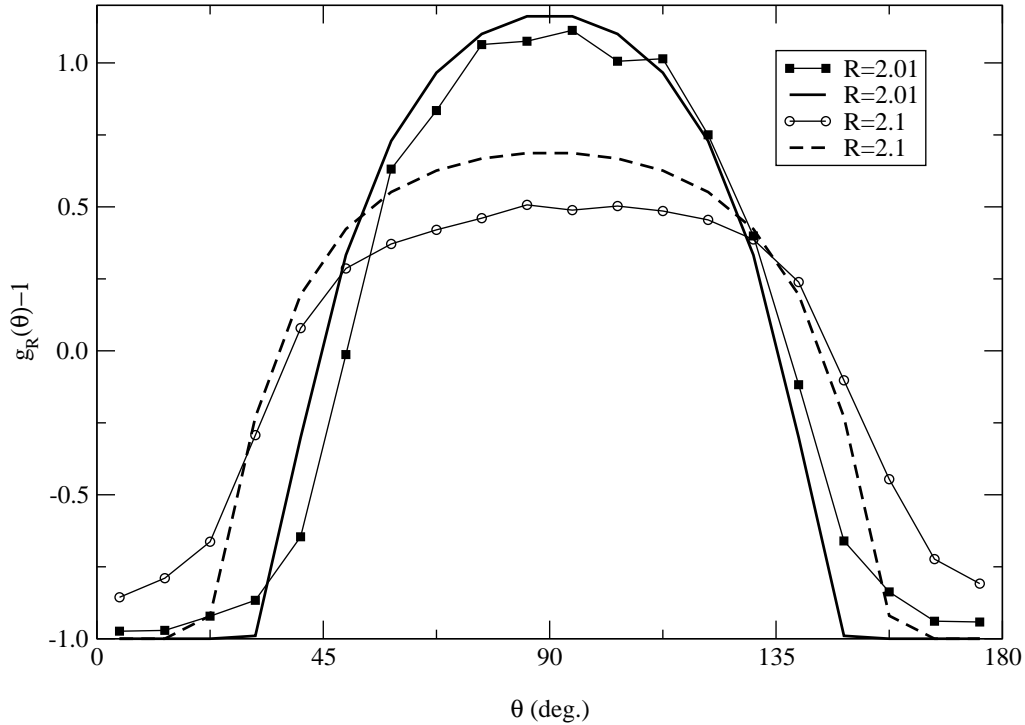}
\caption[]{Angular distribution function $g_R(\theta)$ for pairs of particles corresponding to 
$R=2.01$ and $R=2.1$. The volume fraction is $\phi=5\%$. 
The solid line is obtained using the pair distribution function $g_{\text{\tiny BG}}(r)$ for the region of open
trajectories and assuming zero probability inside the region of closed trajectories.
(see text and figure \ref{closed} for a more detailed explanation).}
\label{gtheta.vs.R}
\end{figure}
It can be seen that, even for this exceedingly short ranged interparticle force, fore-aft symmetry 
is broken at large enough concentrations, and that the distributions show a larger number of pairs 
oriented on the approaching side of the reference sphere than on the receding one. 
More importantly, the angular distribution function seems to reach, in the dilute limit, an asymptotic 
distribution which is anisotropic and shows a depletion of pairs oriented close to the flow direction. This suggests 
a depletion of permanent doublets, which spend more time nearly aligned along the direction of the flow, and seems to 
indicate that, as speculated by \cite{RampallSL97}, any mechanism forcing particles into the region of closed streamlines 
is small compared to the effect of the non-hydrodynamic forces which eliminates particles from this region, and ultimately 
leaves only a negligible number of pairs forming permanent doublets. 
In this case, the pair distribution function in the dilute 
limit should be the combination of $g_{\text{\tiny BG}}(r)$ for the region of open trajectories 
and a zero probability inside the region of closed streamlines.
We show schematically, in figure \ref{closed}, how we can then approximate the
angular dependence of the pair distribution  function of pairs closer than a certain distance $R$, using the
expression for the surface separating the regions of open and closed trajectories,
\cite[][]{batchelor72a},
\begin{equation}
\label{surface}
x_2^2 = \exp \left\{ 2 \! \int^\infty_r \!\! \frac{A(\xi)-B(\xi)}{1-A(\xi)} \frac{d\xi}{\xi} \right\}  
\int^\infty_r \!\!\!\! \frac{B(\xi)}{1-A(\xi)}  
\exp \left\{ -2 \! \int^\infty_\xi \!\! \frac{A(\zeta)-B(\zeta)}{1-A(\zeta)} \frac{d\zeta}{\zeta} \right\}
\xi d\xi.
\end{equation}
Clearly, the surface is axisymmetric with $x_2$ as the symmetry axis.
(Here, and in what follows, the Cartesian axis $1$ lies along the direction of the
mean flow, $2$ is perpendicular to $1$ along the plane of shear, and $3$ is the vorticity axis.) 

In figure \ref{gtheta}, we compare this approximation to the angular dependence of the pair distribution function 
of close pairs ($R=2.01$) with the numerical results, and find a very good agreement for 
concentrations smaller than $5\%$. 
Moreover, in figure \ref{gtheta.vs.R}, we show that this approximation accurately describes the anisotropy found 
for the pair distribution function of pairs with an order of magnitude larger range of separations, i.e. $R=2.1$, 
thus validating the assumption of a complete depletion of pairs forming permanent doublets.

\begin{figure}
\centering
\includegraphics*[width=\Wmax]{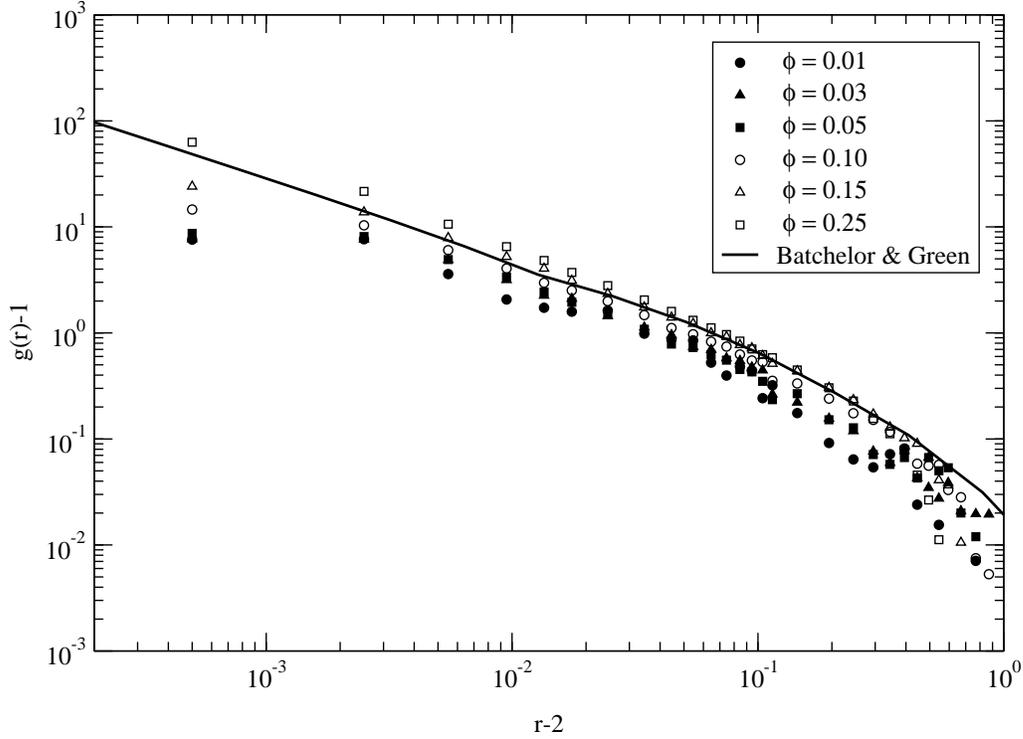}
\caption[]{Pair distribution function integrated over all possible angular orientations. 
Different symbols correspond to the results obtained in numerical simulations for different volume fractions,
$\phi=0.25$, $0.15$, $0.10$, $0.05$ and $0.03$. The pair distribution functions are constructed from a
histogram of the distance between all pairs of particles, averaged over time and over different realizations
(the smallest size of the bins in the histogram is $\Delta r = 0.005$).
The solid line corresponds to the pair distribution function given in Eq. \ref{grbatchelor}.
}
\label{dgr}
\end{figure}
Finally, in figure \ref{dgr}, we present the radial dependence of the pair distribution function, i.e. the pair distribution
function integrated over both spherical angles, as obtained in the numerical simulations at different 
particle concentrations.  
We also compare the numerical results with the pair distribution function given 
in Eq. \ref{grbatchelor}, and find that, although $g_{\text{\tiny BG}}(r)$
does not account for the effect of the closed trajectories, in particular the 
observed depletion of permanent doublets, it both follows the simulation results 
fairly accurately over a wide range of $r$, as well as captures the substantial
increase in the probability of finding pairs of particles near contact.
On the other hand, $g_{\text{\tiny BG}}(r)$ seems to overestimate the asymptotic distribution in 
the dilute limit, and is unable to take into account the effects of the depletion of permanent 
doublets, which would ultimately lead to $g(r)=0$ for $r$ smaller than the minimum possible
separation between approaching spheres in the region of open trajectories, $r_{min} \sim 4\times10^{-5}$. 
But, for distances between the spheres that are not too small, $r \gtrsim 2.001$, Eq. \ref{grbatchelor} 
captures the divergent trend of $g(r)$ as $r \to 2$, which as we shall see, allows us to obtain
a reasonably accurate estimate of the velocity fluctuations in the dilute limit.

\section{Velocity fluctuations}
\label{vf}

Following Batchelor's notation \cite[][]{batchelor72} the temperature tensor 
(the covariance matrix of the velocity fluctuations \cite[][]{vankampen}) 
can be written as,
\begin{equation}
\label{varianceB}
T_{ij} = \frac{1}{N!} \int d {\mathcal C_N} ~ \delta v_i(\boldsymbol r) \delta v_j(\boldsymbol r)  
P(\mathcal C_N|\boldsymbol r),
\end{equation}
where $\delta v_i(\boldsymbol r)$ is the fluctuation in the velocity component $v_i$ for a 
particle located at $\boldsymbol r$ when the configuration of the surrounding spheres is given 
by $\mathcal C_N$, with $P(\mathcal C_N|\boldsymbol r)$ being the probability of such an event. 
From its definition, it is clear that the temperature tensor is symmetric $T_{ij}=T_{ji}$. 
In addition, in simple shear flows there exists an inversion symmetry in the vorticity direction in that, 
a given configuration and its counterpart in which $x_3$ is changed by $-x_3$ are equally probable, 
and therefore we have that $T_{13}=T_{23}=0$, for any volume fraction.
We can simplify the temperature tensor even further by decomposing the simple shear flow into a solid 
body rotating flow, which does not contribute to the velocity fluctuations irrespective of the concentration 
of particles, and a purely straining motion.
The latter is symmetric in $x_1$ and $x_2$ and therefore, 
for any particular velocity fluctuation,  say in the $1$ direction, in a configuration $\cal C_N$ of particles 
surrounding the reference sphere, the same fluctuation but in the $2$ direction would be obtained by a configuration
$\cal C^\prime_N$ in which all the particle positions in $\cal C_N$ are transformed according to $x_1 \leftrightarrow x_2$.
Then, it clearly follows that $T_{11}=T_{22}$, depending only on whether the configurational probability density
$P(\mathcal C_N|\boldsymbol r)$ has the same symmetry, i.e. it is invariant under 
the transformation $x_1 \leftrightarrow x_2$. 
Moreover, since a configuration $\cal C^{\prime \prime}_N$ in which the particle positions in $\cal C_N$ are transformed
according to $x_1 \leftrightarrow -x_1$ would give the negative of the previous velocity fluctuation, and
similarly for fluctuations in the $2$ direction, it is clear that $T_{12}=0$. 
Thus, the temperature tensor should be diagonal, as long as the probability density of particle configurations 
is invariant under those changes, i.e. as long as $P(\mathcal C_N|\boldsymbol r)$ remains invariant under 
the transformations $x_1 \leftrightarrow -x_1$ and $x_2 \leftrightarrow -x_2$. 

In summary, for any concentration and a symmetry preserving configurational probability 
$P(\mathcal C_N|\boldsymbol r)$, we have that the off-diagonals terms of the temperature tensor are null and 
that the temperatures in the plane of shear are equal,
\begin{eqnarray}
T_{ij}&=&0 \qquad i\neq j \\
T_{11}&=&T_{22}
\end{eqnarray}

In the dilute limit, the fluctuations in the velocity come from two-particle interactions, and from the 
far-field form of these interactions it can be shown that any component of the temperature tensor, of the 
form $\delta v_i \delta v_j$, decays faster than $1/r^3$ and therefore, its average value can be directly 
computed by averaging the hydrodynamic interaction between a pair of spheres over all possible configurations,
\begin{equation}
\label{variance}
T_{ij} = \int d {\bf r} ~ \delta v_i \delta v_j \left( 3 \phi / 4 \pi\right) g(\boldsymbol r) = 
\phi  \left[\frac{3}{4 \pi} \int d {\bf r} ~ \delta v_i \delta v_j g(\boldsymbol r) \right] =
\phi ~ t_{ij}, 
\end{equation}
which gives a linear dependence of the temperature components on the volume fraction, 
$T_{ij} = \phi~t_{ij}$.

For two freely-moving spheres in a simple shear flow, 
the velocity fluctuation of a sphere induced by a second sphere the center of 
which is located at $\boldsymbol r$ is given by \cite[][]{daCunhaH96}:
\begin{eqnarray}
\label{vel_fluc1}
\delta v_1 &=& \dot{x_1}-x_2 = e x_1 - \frac{1}{2} B x_2 \\
\label{vel_fluc2}
\delta v_2 &=& e x_2 - \frac{1}{2} B x_1 \\
\label{vel_fluc3}
\delta v_3 &=& e x_3,
\end{eqnarray}
where $e=x_1 x_2 (B-A)/r^2$. Using these equations, it can be easily shown that for a pair distribution function 
which depends only on $r$, the temperature tensor is not only diagonal, but that $T_{33}$, its component in the 
vorticity direction, is smaller than the temperature in the plane of shear,
\begin{equation}
T_{11} = T_{22} > T_{33}.
\end{equation}

\begin{table}
\begin{center}
\begin{tabular}{|l|c|c||l|c|c|}
\cline{1-6}
&  $t_{11}=t_{22}$&$t_{33}$ &
&  $t_{11}=t_{22}$&$t_{33}$ \\
\cline{1-6}
Random Hard Sphere ($g_{\text{\tiny HS}}(r)$)	& 0.3157 & 0.0811 &
Simple Shear Flow  ($g_{\text{\tiny BG}}(r)$)		& 0.4637 & 0.1031 \\ 
~~~~~~~~~ lubrication 	& 0.0040 & 0.0006 &
~~~~~~ lubrication 	& 0.0896 & 0.0117 \\
~~~~~~~~~ intermediate 	& 0.0930 & 0.0175 &
~~~~~~ intermediate 	& 0.1531 & 0.0279 \\
~~~~~~~~~ far-field 	& 0.2187 & 0.0630 &
~~~~~~ far-field 	& 0.2210 & 0.0635 \\
\cline{1-6}
\end{tabular}
\end{center}
\caption{Temperature tensor in the dilute limit, computed using Eqs.~\ref{variance}-\ref{vel_fluc3} 
for two different pair distribution functions, one for a hard-sphere distribution, 
and the other given by \cite{batchelor72b} for a simple shear flow in the dilute limit, Eq.~\ref{grbatchelor}. 
Also given is the contribution to the velocity fluctuations coming from the 
three different regions in which the mobility functions $A$ and $B$ are divided, i.e. the lubrication 
region ($2 < r \leq 2.01$), the intermediate region ($2.01 < r \leq 2.5$) and the far-field region 
($2.5 < r$).}
\label{vf_theo}
\end{table}
The exact temperature values will depend in general on the pair distribution function. In Table \ref{vf_theo}, 
we present the diagonal terms of the temperature tensor in the dilute limit, obtained from the numerical 
integration of Eq. \ref{variance} for two different isotropic pair distribution functions, corresponding to a 
random distribution of hard spheres, $g_{\text{\tiny HS}}(r)=1$,  
and to Batchelor \& Green's result for a suspension in a simple shear flow, $g_{\text{\tiny BG}}(r)$
given by Eq.~\ref{grbatchelor}  \cite[][]{batchelor72b}. 
Note that, although in our numerical simulations we observe a depletion of permanent doublets
in the dilute limit, we first neglect this effect, and compute the temperature terms by numerically
calculating the integrals in Eq.\ref{variance} using $g_{\text{\tiny BG}}(r)$, 
for all possible angular orientations. 
As we shall see, for dilute suspensions, this provides a satisfactory
approximation to the temperature tensor computed from our numerical simulations. 
The contribution to the integral of each region in which the mobility functions
$A$ and $B$ are divided, i.e. the lubrication, intermediate and far-field regions, 
is also given in Table 
\ref{vf_theo}. 
As expected, the far-field contribution is practically identical in both cases, since in fact 
 $g_{\text{\tiny BG}}(r)$ asymptotically approaches $g_{\text{\tiny HS}}(r)$ for $r \to \infty$. 
On the other hand, the contribution of the lubrication region to the velocity fluctuations is at least an order 
of magnitude larger if computed using $g_{\text{\tiny BG}}(r)$ because, in that case, the 
probability of finding two nearly touching spheres is substantially larger than in the hard-sphere case. 
On the other hand, the anisotropy ratio in the dilute limit is similar in both cases, $T_{11}/T_{33} \sim 4$.

\begin{figure}
\centering
\includegraphics*[width=\Wmax]{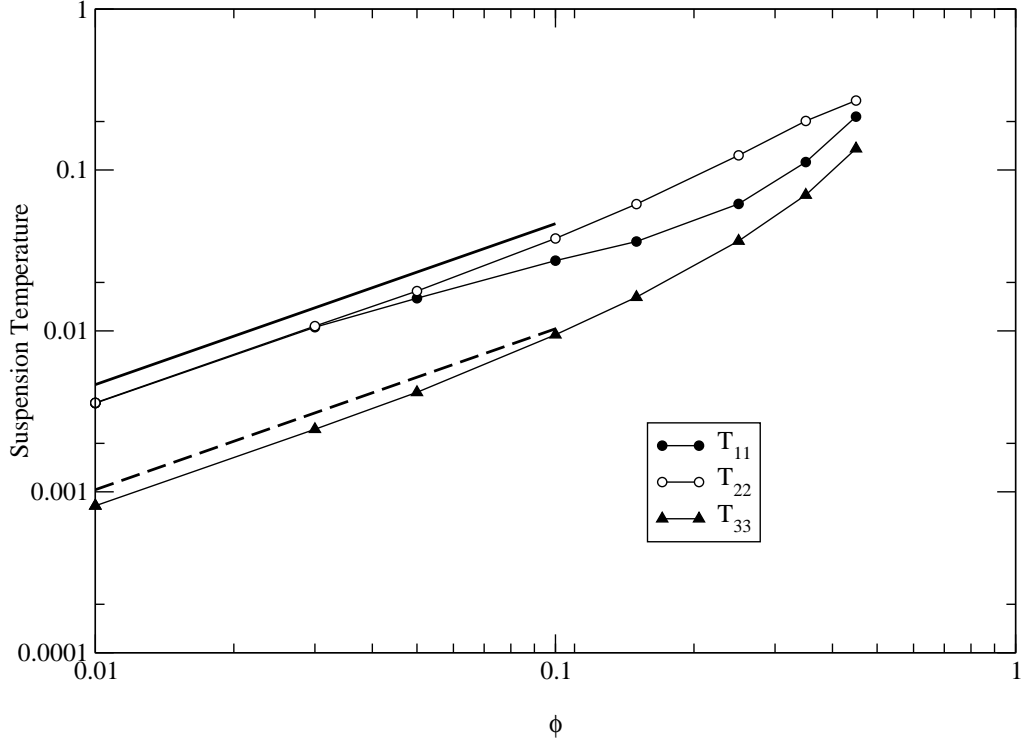}
\caption[]{Diagonal components of the temperature tensor as a function of the volume fraction, obtained from the numerical Stokesian dynamics simulations. The solid and dashed lines correspond to the dilute
limit calculation for $T_{11;22}=\phi ~t_{11;22}$ and $T_{33}=\phi ~t_{33}$, respectively.
The computed values of $t_{11;22}$ and $t_{33}$ are given in the second part of table \ref{vf_theo}.
The discrepancy between theory and numerical results is about 25\%.}
\label{temp_diag}
\end{figure}

In figure \ref{temp_diag}, we present the diagonal terms of the temperature tensor as a function of the volume 
fraction, obtained in a simple shear flow by means of Stokesian dynamics simulations. 
The temperature components $T_{11}$ and $T_{22}$ converge to a common curve in the dilute limit, 
which is consistent with the existence of an isotropic pair distribution function and indicates that the effect of 
the particle-depleted region of closed streamlines is not 
measurable. 
In addition, the decay of the velocity fluctuations follows the dilute limit scaling given by 
Eq.~\ref{variance}, viz. that $T_{ij}$ is proportional to $\phi$, even for surprisingly high 
volume fractions. 
On the other hand, at larger concentrations we see that the $T_{11}$ and $T_{22}$ curves separate from each other,
which is evidence of the structure developed by the suspension at high concentrations. 
In fact, in figure \ref{gtheta} we showed that, although we have a very short-ranged interparticle force, 
$r_c=10^{-4}$, the pair distribution function has fore-aft symmetry in the dilute limit,
for larger concentrations this symmetry no longer holds.  
\begin{figure}
\centering
\includegraphics*[width=\Wmax]{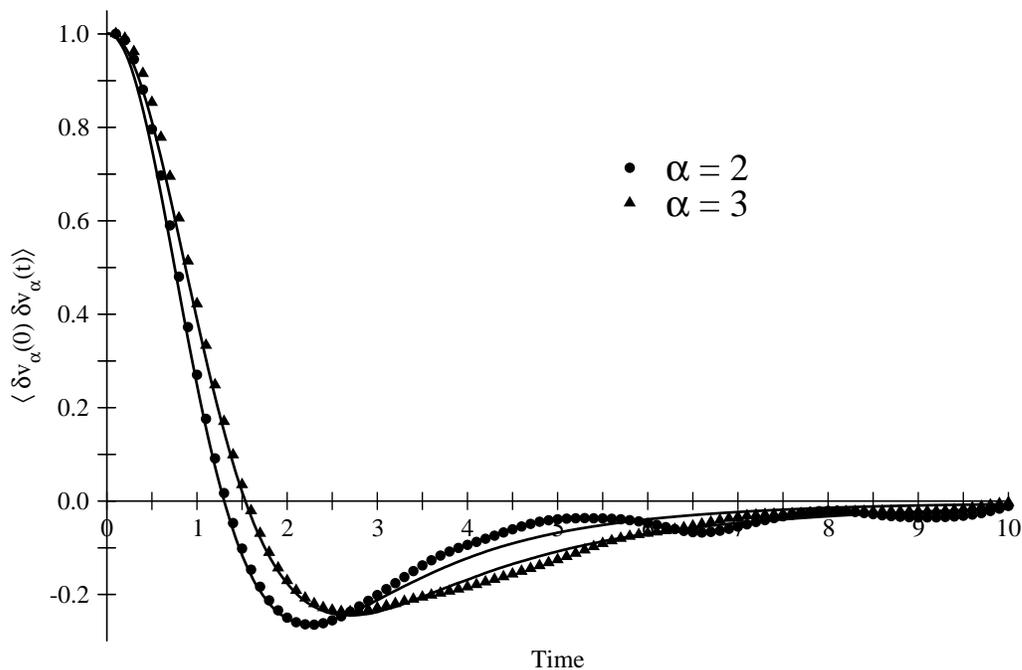}
\caption[]{Velocity autocorrelation function in the dilute limit in both transverse directions. 
Symbols correspond to the results obtained using the SD method for a volume fraction $\phi=0.03$. 
Solid lines correspond to the numerical computation of the velocity autocorrelation function using two-particle
interactions only.}
\label{autocorrela}
\end{figure}

The agreement observed in figure \ref{temp_diag}, between the suspension temperature computed in the dilute limit
using $g_{\text{\tiny BG}}(r)$ and via the numerical simulations, shows that, as expected, 
the dominant contribution to the velocity fluctuations arises from two-particle hydrodynamic interactions.
Furthermore, in paper I, we argued that in the dilute limit, the whole velocity autocorrelation function converges 
to an asymptotic function dominated by two-particle encounters. 
To further investigate the 
dilute limit behavior of the velocity fluctuations, we therefore computed the velocity autocorrelation function
on the basis only o two-particle hydrodynamic interactions, using Eqs. \ref{vel_fluc1}-\ref{vel_fluc3}.
In order to do that, we simulated a large number
of two-particle encounters ($N_c=2\times10^5$) between a test sphere, initially located at the origin, and
an incoming particle, initially far away from the test sphere.
The exact position of the incoming particle was chosen randomly in the region 
($-\Delta x_1 < x^0_1 < -\Delta x_1 + R$, $0< x^0_2<R$, $0<x^0_3<R$), 
where $R$ is the size of the cross-section for the incoming particles  considered in the calculations. 
Note that, in order for an encounter to induce a significant velocity fluctuation in the test sphere, 
both spheres should come reasonably close to each other at some point during their motion, and we thus 
use $R=5$ \cite[][]{wang96}($\Delta x_1$ was set to $20R$).
The probability distribution used to generate the initial conditions was uniform in $x_1$ and $x_3$,
and in the shear direction we used a probability distribution proportional to the incoming flux of
particles in simple shear flow, that is $p(x_2)\propto x_2$. 
The cross-section of the region of closed streamlines, perpendicular to the flow direction, 
is so small at $\Delta x_1=20R$, that we did not observe any closed trajectory after simulating 
$N_c$ encounters.
The motion of both spheres was then computed, using Eqs.\ref{vel_fluc1}-\ref{vel_fluc3} and a time step
$\Delta t=10^{-5}$, until the incoming particle reached the point that was symmetric with respect to
its initial position, that is ($-x^0_1$, $x^0_2$, $x^0_3$). 
Finally, the velocity autocorrelation function was computed by
averaging over all the simulated trajectories, after splitting each one of them into 
intervals of time $T=10$. 
The results are shown in figure \ref{autocorrela}, where the numerically
computed velocity autocorrelation functions in both transverse directions is compared to the
results obtained by means of Stokesian dynamics simulations (already presented in paper I). 
An excellent agreement is obtained, which confirms that the velocity autocorrelation functions in both
transverse directions reach corresponding asymptotic forms in the dilute limit.
It also demonstrates that, as was first suggested in paper I, the fact that
both autocorrelation functions become negative at times $t\sim1$ is due to the 
anti-correlated motion performed by the spheres during binary collisions, i.e. the transverse
velocities of the spheres involved in a binary collision 
are reversed at the instant at which the incoming particle
goes from the approaching to the receding side of the reference sphere.

\begin{figure}
\centering
\includegraphics*[width=\Wmax]{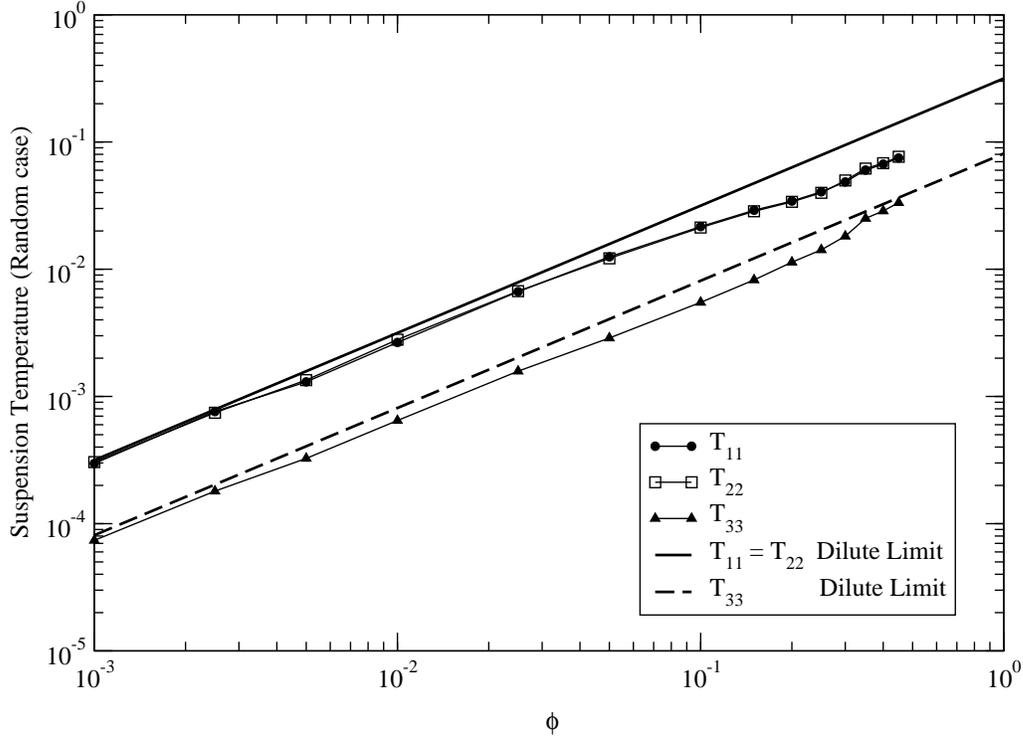}
\caption[]{Diagonal components of the temperature tensor as a function of the volume fraction, as 
obtained for a random distribution of hard spheres. We assume that the spheres are subject to a simple
shear flow and compute their velocity using the Stokesian dynamics code. 
The dashed and dotted lines correspond to the dilute
limit calculation $T_{ij}=\phi ~t_{ij}$ and $t_{ij}$ given in the first part of 
table \ref{vf_theo}.}
\label{temp_diag_random}
\end{figure}

In order to investigate the effects of the steady state microstructure developed by the sheared suspensions, 
we performed a second type of numerical computations, in which the velocity fluctuations were calculated 
for a random hard-sphere distribution of particles subject to simple shear flow. 
That is, we first generated a random distribution of spheres at the desired concentration, and then, using the Stokesian
dynamics code, we calculated the instantaneous velocity of all the spheres in the presence of a simple shear flow.
We then averaged the results over many different realizations of the random hard-sphere distribution. 
Typically, the number of particles in each realization was the same as in the dynamic simulations, 
but the number of configurations
was increased to $N_c=256$. 
We refer to the randomly generated hard-sphere particle distribution as {\it Hard Sphere} (HS) distribution
in contrast to the {\it Shear Flow} (SF) distribution which refers to the particle distribution 
which is attained asymptotically after the suspension has been sheared in a simple shear flow for strains 
in excess of $50$.
\begin{figure}
\centering
\includegraphics*[width=\Wmax]{temp_off_diag.eps}
\caption[]{Off-diagonal terms of the temperature tensor as a function of the volume fraction}
\label{temp_off_diag}
\end{figure}
\begin{figure}
\centering
\includegraphics*[width=\Wmax]{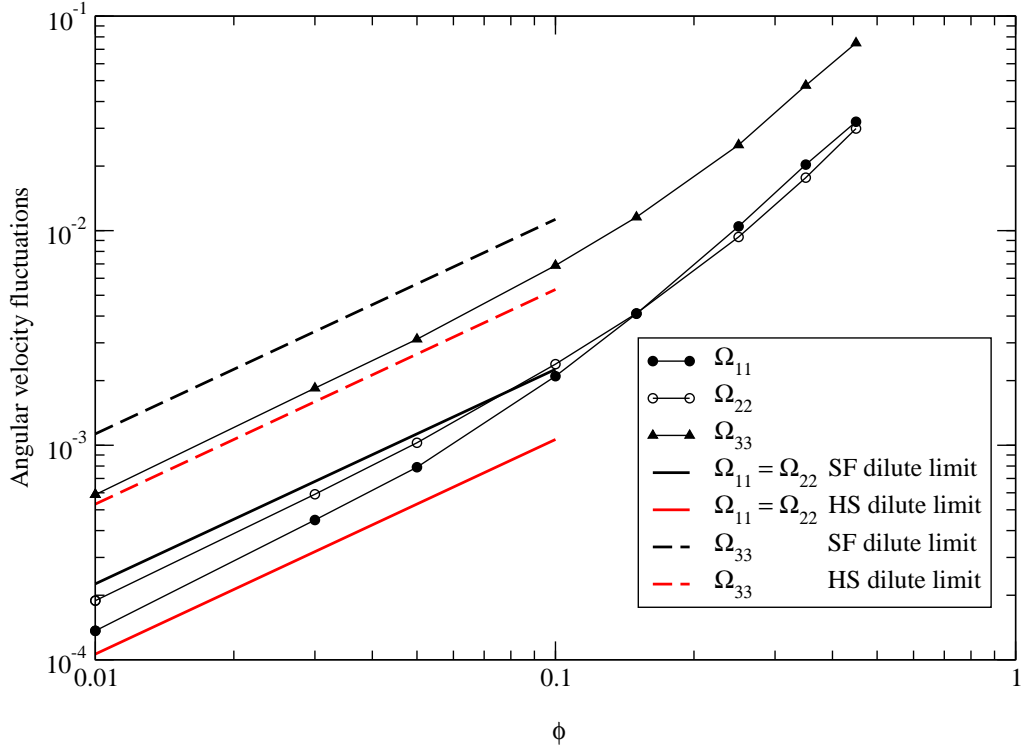}
\caption[]{Diagonal components of the angular velocity autocorrelation tensor as a function of the volume fraction, 
obtained from the Stokesian dynamics simulations. The solid and dashed lines correspond to the dilute
limit calculation for $\Omega_{11;22}=\phi ~\tilde\Omega_{11;22}$ and $\Omega_{33}=\phi ~\tilde\Omega_{33}$, 
respectively. Here the black and gray lines correspond to the dilute limit calculations using
SF and HS distributions respectively, and the corresponding values of $\tilde\Omega_{11;22}$ and $\tilde\Omega_{33}$ 
are given in table \ref{wf_theo}.
The discrepancy between theory and numerical results is close to a factor $2$.
The results are bounded between the calculations using SF and HS distributions.} 
\label{tempW_diag}
\end{figure}

In figure \ref{temp_diag_random}, we plot the diagonal components of the temperature tensor, obtained 
for HS distributions in simple shear flow. We can see that, as a result of the isotropic spatial distribution 
of particles, $T_{11}$ equals $T_{22}$ for all values of the volume fraction (within 3\%). Also, $T_{33}$,
the temperature in the vorticity direction is smaller than that in the plane of shear, as predicted
in the dilute limit.
The agreement between theory and numerical results is excellent in this case, with a discrepancy
less than 10\% for the lowest volume fraction.

As mentioned earlier, the off-diagonal terms of the temperature tensor are expected to be zero 
for an isotropic pair distribution function. 
Moreover, if the only broken symmetry is the fore-aft symmetry, the only term
that may differ from zero is $T_{12}$, which should actually be negative if particles are depleted 
in the receding side of the reference sphere, as is observed in the numerical and experimental work. 
The numerical results agree completely with this analysis, as shown in figure \ref{temp_off_diag}
 in that, for all volume fractions, all the off-diagonal components vanish for the HS distributions, 
as well as $T_{13}$ and $T_{23}$ for
the SF distributions. 
Moreover, the only correlation present in the velocity fluctuations is given
by $T_{12}$, which becomes different from zero only as the concentration increases and a fore-aft
asymmetry is developed by the SF distributions.
\begin{table}
\begin{center}
\begin{tabular}{|cc|c||cc|c|}
\cline{1-6}
& & $\tilde\Omega_{11}=\tilde\Omega_{22}=\tilde\Omega_{33}/5$ && & $\tilde\Omega_{11}=\tilde\Omega_{22}=\tilde\Omega_{33}/5$ \\
\cline{1-6}
\cline{1-6}
\multicolumn{2}{|l|}{Random Hard Sphere ($g_{\text{\tiny HS}}(r)$)} & 0.01064 &
\multicolumn{2}{|l|}{Simple Shear Flow  ($g_{\text{\tiny BG}}(r)$)} & 0.02260    \\
\cline{1-6}
\cline{1-6}
& ~~~~~lubrication & 0.00033 &
& ~~~lubrication~~~ & 0.00869 \\
& ~~ intermediate & 0.004438 &
& intermediate & 0.007915\\
& ~~~~~far-field & 0.00587 &
& ~~~far-field~~~ & 0.00600 \\
\cline{1-6}
\end{tabular}
\end{center}
\caption{Angular velocity fluctuations in the dilute limit, computed for two different assumed pair 
distribution functions, one corresponding to a random distribution of hard spheres, and the other 
using $g_{\text{\tiny BG}}(r)$.
The contribution to the velocity fluctuations coming from the three different regions in 
$g_{\text{\tiny BG}}(r)$ and $C(r)$ - c.f. Eq. \ref{cr} - are noted separately, 
i.e. the lubrication region ($2 < r \leq 2.01$), the intermediate
region ($2.01 < r \leq 2.5$) and the far-field region ($2.5 < r$).}
\label{wf_theo}
\end{table}

A completely analogous analysis to the one presented above for the linear velocity fluctuations 
gives very similar results for the fluctuations in the angular velocity, 
$\Omega_{ij} = \langle \delta \omega_i \delta \omega_j \rangle$. 
The off-diagonal terms are zero ($\Omega_{ij}=0, ~i \neq j$) and $\Omega_{11}=\Omega_{22}$, for all
concentrations, as long as the distribution of spheres has the symmetries discussed previously
when we analyzed the properties of the temperature tensor. 

In the dilute limit, the fluctuations can be written as,
\begin{equation}
\label{Wvariance}
\Omega_{ij} = \int d {\bf r} ~ \delta \omega_i \delta \omega_j \left( 3 \phi / 4 \pi\right) g(\boldsymbol r) = 
\phi  \left[ \frac{3}{4 \pi}\int d {\bf r} ~ \delta \omega_i \delta \omega_j g(\boldsymbol r) \right] =
\phi ~ \tilde\Omega_{ij}. 
\end{equation}
and for two freely-moving spheres we have that \cite[][]{batchelor72a}:
\begin{eqnarray}
\label{ang_fluc}
\delta \omega_1 &=& - \frac{1}{2} C \frac{x_1~x_3}{r^2}, \\
\label{ang_fluc2}
\delta \omega_2 &=& - \frac{1}{2} C \frac{x_2~x_3}{r^2}, \\
\label{ang_fluc3}
\delta \omega_3 &=& + \frac{1}{2} C \frac{x_1^2-x_2^2}{r^2},
\end{eqnarray}
where $C$ is a dimensionless function of $r$ only. 
Using Eqs. \ref{Wvariance}-\ref{ang_fluc3}, and assuming an isotropic pair distribution function,
it can be shown that, $\Omega_{33}=5~\Omega_{11}=5~\Omega_{22}$
for any radial dependence of the pair distribution function. 
Finally, using the far-field and the lubrication approximations of $C(r)$ given
by \cite{kim}, and a linear interpolation in the intermediate region, 
\begin{equation}
\label{cr}
C(r) = \left\{     
\begin{array}{llc}
C_l(r)=\frac{-2.283 + 2.3052 L + 0.2972 L^2}{6.32549 + 6.0425 L + L^2}
& \rm{for} & r < 2.01 \\
C_i(r) = \frac{2.5-r}{2.5-2.01} C_l(r) + \frac{r-2.01}{2.5-2.01} C_f(r)
& \rm{for} & 2.01 < r < 2.5 \\
C_f(r)=\frac{5}{2 r^3} - \frac{25}{4 r^6} + \frac{125}{8 r^9} + \frac{25}{r^{10}} + \frac{125}{2 r^{11}}   
& \rm{for} & 2.5 < r
\end{array}
\right.,
\end{equation}
we obtain the results presented in table \ref{wf_theo} for $\tilde\Omega_{ij}$.
\begin{figure}
\centering
\includegraphics*[width=\Wmax]{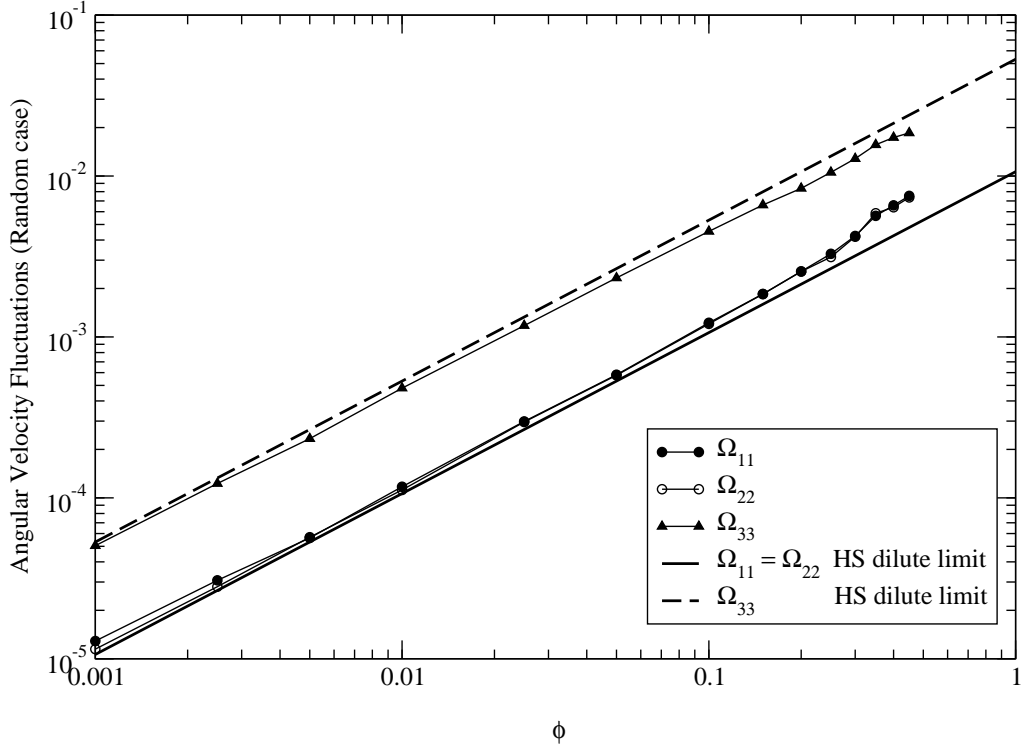}
\caption[]{Diagonal components of the angular velocity fluctuations as a function of the volume fraction, 
obtained for HS distributions, subject to a simple shear flow, and computed using the Stokesian dynamics code. 
The solid and dashed lines correspond to the dilute
limit calculation $\Omega_{ij}=\phi ~\tilde\Omega_{ij}$ with $\tilde\Omega_{ij}$ being given in the first part of 
table \ref{wf_theo}. The difference between theory and numerical simulations is less than $20\%$ at
the lowest concentrations.}
\label{tempW_diag_random}
\end{figure}

In figures \ref{tempW_diag} and \ref{tempW_diag_random}, we present the velocity fluctuations in the 
angular velocity obtained for the SF and the HS distributions of particles. 
These fluctuations seem to be very sensitive to the
microstructure developed by the suspension because, 
even at very low concentrations, a difference between
$\tilde\Omega_{11}$ and $\tilde\Omega_{22}$ can be observed in the SF case, contrary to what 
happens in the HS case where they coincide for all concentrations, as expected from our
previous discussion. 
Also, the discrepancy between the theoretical and the numerical values is large for the SF distribution 
(a factor $\sim 2$), while, in the HS case, there is good agreement with the theory. 
Not even the linear behavior on $\phi$ in the dilute limit seems to have been reached in the SF case
at low concentrations, in contrast to what is observed for the HS distributions of spheres. 
However, we also show in figure \ref{tempW_diag} that, the calculations using $g_{\text{\tiny HS}}(r)$
provide a lower bound to the angular velocity fluctuations. Qualitatively, this behavior can be understood
from the observation that the main difference between the results obtained for HS and SF distributions comes
from the contribution of the lubrication region to the fluctuations, which is negligible in the HS case. 
Thus, a lower bound to the velocity fluctuations in the absence of permanent doublets can be estimated roughly
using the HS distributions, which have a negligible contribution from the lubrication region. 

\begin{figure}
\centering
\includegraphics*[width=\Wmax]{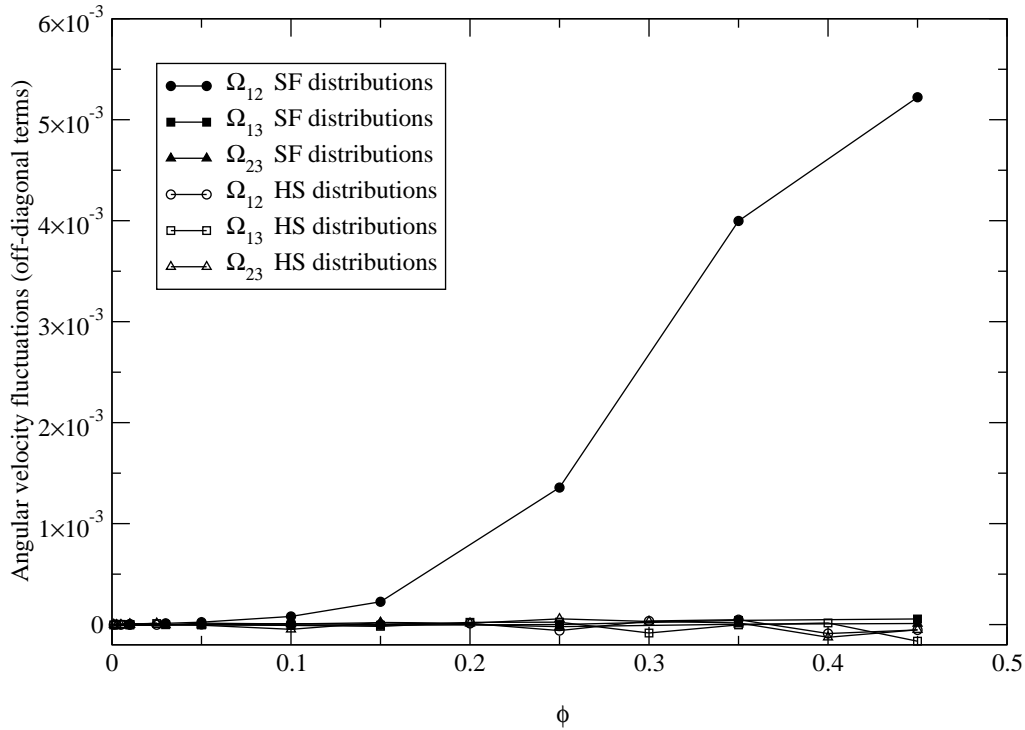}
\caption[]{Off-diagonal terms of the angular velocity autocorrelation 
tensor as a function of the volume fraction}
\label{tempW_off_diag}
\end{figure}

Finally, we present the off-diagonal terms of the angular
velocity autocorrelation tensor. 
As predicted, all terms are negligible in the dilute limit, and
only the microstructure developed by the SF distributions leads to a correlation $\tilde\Omega_{12} \neq 0$,
which in fact is in agreement with the result, referred to earlier, that $t_{12} \neq 0$, 
and with the observed depletion of pairs of particles oriented on the receding side of the interaction
(see figure \ref{gtheta}).

\begin{figure}
\centering
\includegraphics*[width=\Wmax]{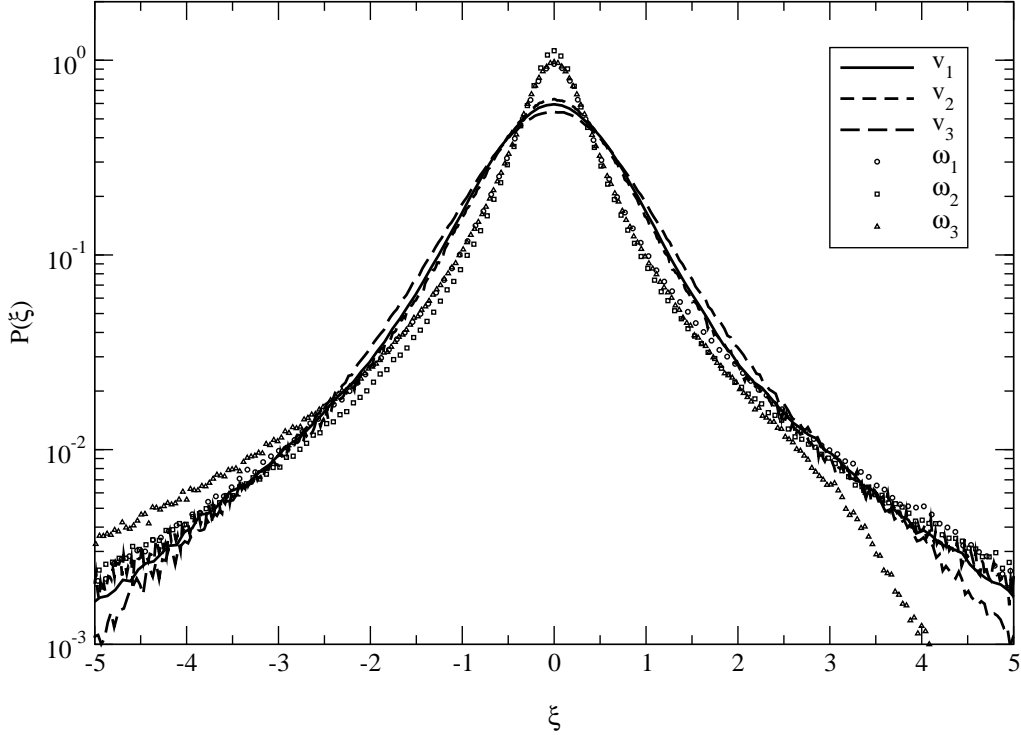}
\caption[]{Probability density functions for the three components of the linear velocity fluctuations and 
the three components of the angular velocity fluctuations, at very low concentrations $\phi=0.01$. 
All p.d.f.'s for the indicated variables, $\delta v_1, v_2, v_3, \omega_1, \omega_2$ and $\delta \omega_3$,
are rescaled using the corresponding temperatures so that the standard deviation is unity 
(for example $\xi=\delta v_1/\sqrt{T_{11}}$, and similarly for the other variables).}
\label{pdf_phi01}
\end{figure}

\begin{figure}
\centering
\includegraphics*[width=\Wmax]{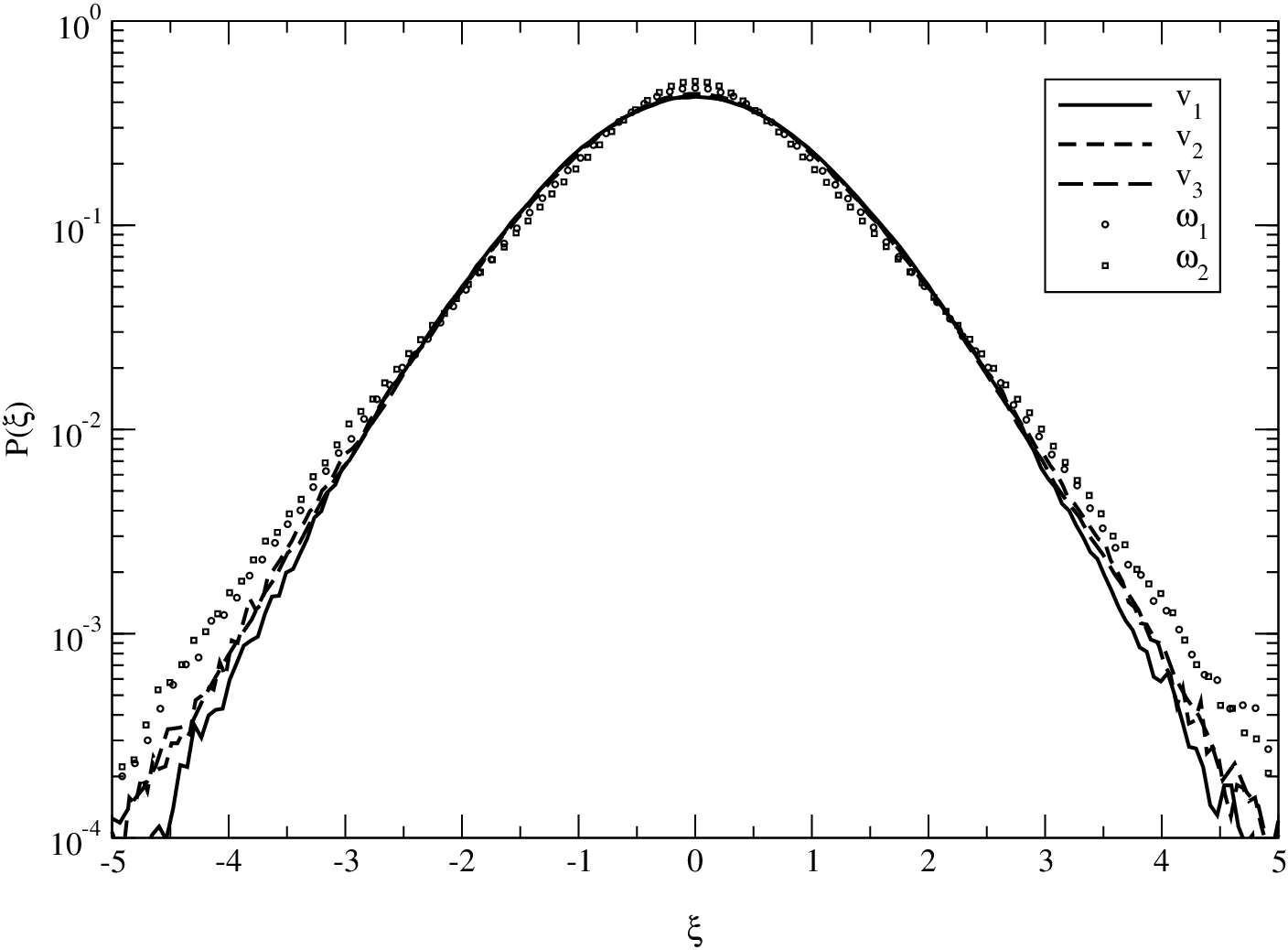}
\caption[]{Probability density functions for the three components of the linear velocity fluctuations and 
the two components of the angular velocity fluctuations in the plane of shear. The vorticity component of the
angular velocity fluctuations is not symmetric as it can be observed in figure \ref{wz}.
 The volume fraction is  $\phi=0.10$.  
All p.d.f.'s for the indicated variables, $\delta v_1, v_2, v_3, \omega_1, \omega_2$ and $\delta \omega_3$,
are rescaled using the corresponding temperatures so that the standard deviation is unity 
(for example $\xi=\delta v_1/\sqrt{T_{11}}$, and similarly for the other variables).}
\label{pdf_phi10}
\end{figure}

\begin{figure}
\centering
\includegraphics*[width=\Wmax]{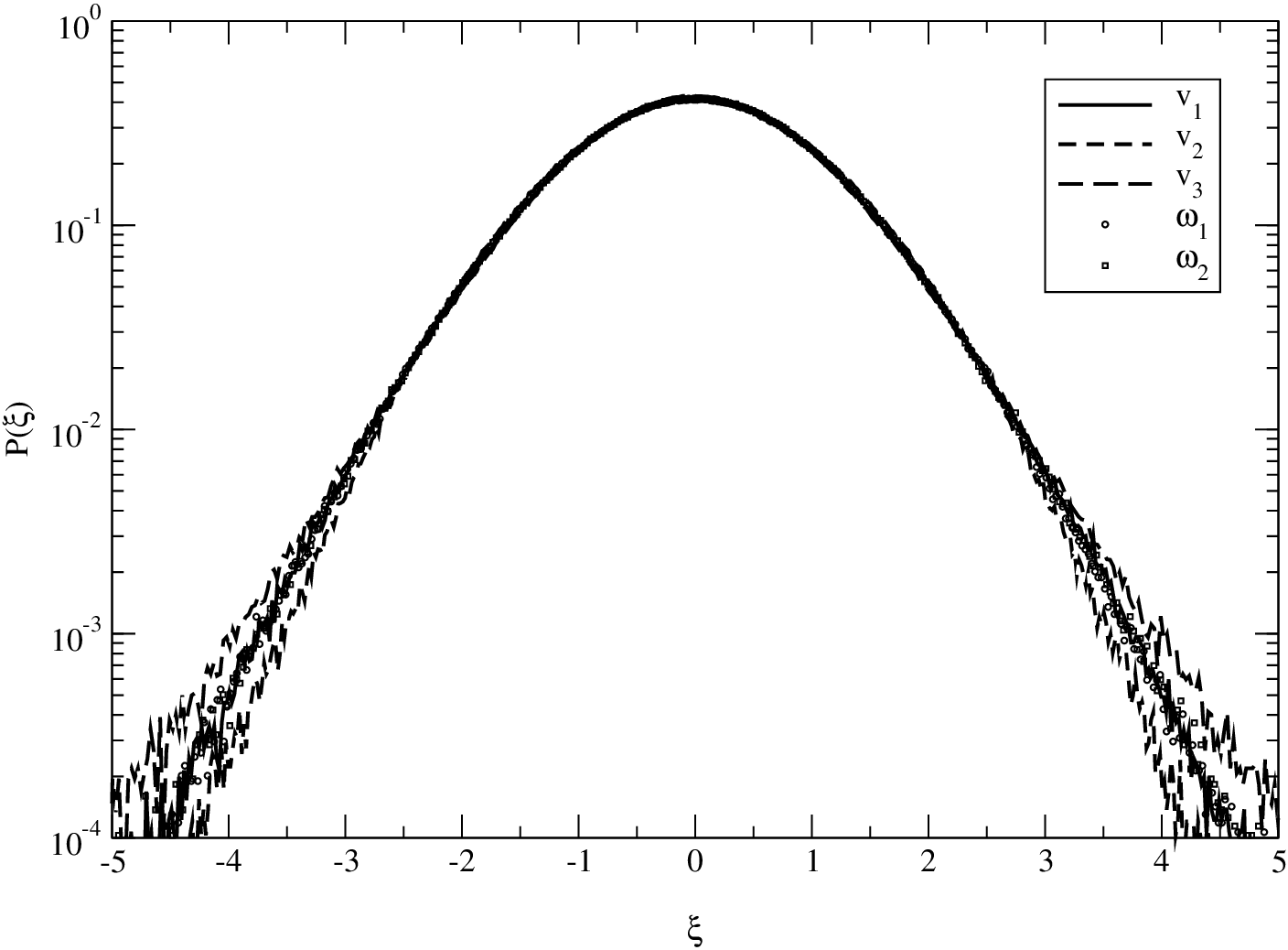}
\caption[]{Probability density functions for the three components of the linear velocity fluctuations and 
the two components of the angular velocity fluctuations in the plane of shear. The vorticity component of the
angular velocity fluctuations is not symmetric as it can be observed in figure \ref{wz}.
 The volume fraction is  $\phi=0.45$.  
All p.d.f.'s for the indicated variables, $\delta v_1, v_2, v_3, \omega_1, \omega_2$ and $\delta \omega_3$,
are rescaled using the corresponding temperatures so that the standard deviation is unity 
(for example $\xi=\delta v_1/\sqrt{T_{11}}$, and similarly for the other variables).}
\label{pdf_phi45}
\end{figure}
\begin{table}
\begin{center}
\begin{tabular}{|c|c|c|c|c||c|c|c|c|c||c|c|c|c|c|}
\cline{1-15}
\multicolumn{15}{|c|}{$P(\xi) \propto e^{-\alpha \xi^{\beta}}$}\\
\cline{1-15}
$\phi$& $v_i$ & $\alpha$ & $\omega_i$& $\alpha$ &
$\phi$& $v_i$ & $\alpha$ & $\omega_i$& $\alpha$ &
$\phi$& $v_i$ & $\alpha$ & $\omega_i$& $\alpha$ \\
\cline{1-15}
0.45 &$v_1$ & 2.54 & $\omega_1$ & 16.1 & 
0.10 &$v_1$ & 14.1 & $\omega_1$ & 39.2 & 
0.01 &$v_1$ & 18.1 & $\omega_1$ & 22.8 \\
\cline{2-5} \cline{7-10} \cline{12-15}
$\beta=2$ &$v_2$ & 1.85 & $\omega_2$ & 17.8 & 
$\beta=1$     &$v_2$ & 10.9 & $\omega_2$ & 36.0 &
$\beta=1/4$     &$v_2$ & 18.1 & $\omega_2$ & 21.7 \\
\cline{2-5} \cline{7-10} \cline{12-15}
     &$v_3$ & 4.54 &  & &
     &$v_3$ & 21.1 &  & &
     &$v_3$ & 24.8 &  &   \\
\cline{1-15}
\end{tabular}
\end{center}
\caption{Fitted values for the probability density function of all the linear and angular
velocity components.}
\label{alfa}
\end{table}

\begin{figure}
\centering
\includegraphics*[width=\Wmax]{pwz.eps}
\caption[]{Probability density functions of the angular velocity in the vorticity direction, $P(\omega_3)$, 
for five different volume fractions $\phi=0.45$, $0.35$, $0.25$, $0.15$ and $0.05$.} 
\label{pwz}
\end{figure}

\begin{figure}
\centering
\includegraphics*[width=\Wmax]{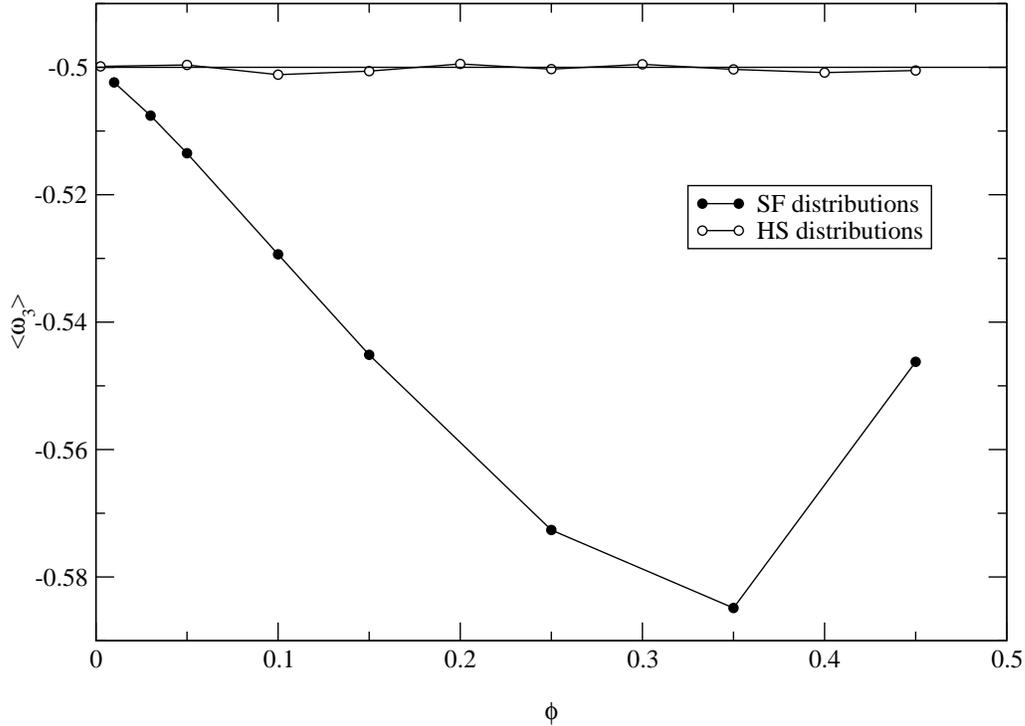}
\caption[]{Mean value of the angular velocity in the vorticity direction 
$\langle \omega_3 \rangle$ as a function of the volume fraction $\phi$ of the suspension. Open symbols
corresponds to the same mean angular velocity but for the HS distributions, 
calculated using the Stokesian dynamics code.}
\label{wz}
\end{figure}

In addition to the temperature values, the numerical simulations provide us with greater detail about 
the velocity fluctuations. In fact, we can obtain the full probability distribution function for the
fluctuations in both the linear and angular velocity components, calculated from a histogram of the 
particle velocities, averaged both over $N_c$ different realizations and in time.

In figures \ref{pdf_phi01}, \ref{pdf_phi10} and \ref{pdf_phi45}, we show the normalized p.d.f of 
the linear and angular velocity fluctuations in all directions, and for three different volume 
fractions. (Note that, for the velocity component in the direction of the shear, we have subtracted the
contribution of the external velocity field at the center of the particle, that is $\delta v_1=\dot x_1 - x_2$.) 
Different functional forms are observed as the concentration decreases. 
A first transition, from Gaussian to Exponential distributions occurs when the 
concentration is decreased from $\phi=0.45$ to $\phi=0.10$, as already presented 
in paper I. 
A second transition, from Exponential to a {\it Stretched} Exponential
with exponent $\sim 1/4$ occurs when the concentration decreases even further down 
to $\phi=0.01$. 
All the numerical data were fitted using exponential distributions of the form 
$P(\xi) \propto exp(-\xi^\alpha)$, with $\alpha = 2$ for large concentrations 
($\phi=0.45$), $\alpha=1$ for intermediate values of the volume fraction ($\phi=0.10$),
and $\alpha=1/4$ for very low concentrations ($\phi=0.01$). 
In paper I, we discussed 
the first transition, from Gaussian to Exponential distributions, by analogy with 
turbulent flows, where one observes this type of transition in the p.d.f of the velocity
differences, the temperature and other passive scalars. The second transition presented
here is in accordance with that analogy, in that, with decreasing concentrations,
the exponent $\alpha$ of the distribution also decreases, implying {\it intermittency},
in that the probability of rare events is much larger than expected from Gaussian
statistics \cite[][]{sreenivasan99}.

Finally, the vorticity component of the angular velocity, $\omega_3$, has a mean value different from zero, due 
to the shear flow. In the Stokes limit, by decomposing the linear shear into a purely rotational flow and a
purely straining flow, it is easy to show that the angular velocity of a single
sphere is $\Omega_3 = -1/2$ \cite[][]{leal}, which is therefore the expected average value
in the dilute limit. For larger concentrations, however, hydrodynamic interactions between spheres should be
taken into account. But, using the same superposition of flows, and due to the reflection 
symmetry of the purely straining flow, it can be shown that the average remains constant, and 
equal to $\Omega_3$, even in the presence of other spheres, as long as the distribution of spheres is 
isotropic. 
In figure \ref{pwz}, we show the distribution
of angular velocities in the vorticity direction, for $\phi=0.45$, $0.35$, $0.25$, $0.15$ and $0.05$,
while in figure \ref{wz},
we show the mean angular velocity as a function of the volume fraction. As can be seen, $\langle \omega_3 \rangle$   
decreases from $-1/2$ down to $-0.58$ at $\phi=0.35$ and then
increases to $-0.55$ at $\phi=0.45$. (However, note that the shift in 
$\langle \omega_3 \rangle$ with respect to $\Omega_3=-1/2$ is always smaller than the
width of the distribution $\sigma_{\omega_3}$, and therefore that the fluctuations are larger than the shift in
the average value.) 
We also show the results obtained
for the HS distributions, calculating $\langle \omega_3 \rangle$ using Stokesian dynamics, 
where it can be seen that the mean angular velocity remains equal to $\Omega_3=-1/2$
for all concentrations. 
We can conclude therefore, that the deviation in the mean angular velocity
is due to the anisotropy developed by the suspension in simple shear flows. 
Moreover, as shown in figure 3 of paper I, the angular
dependence of the pair distribution function for close spheres shows a larger probability
for pairs oriented at angles $45^\circ < \theta < 135^\circ$, which is consistent with an increase
in the angular speed of the spheres.

\subsection{A note on LDV measurements}

\begin{figure}
\centering
\includegraphics*[width=\Wmax]{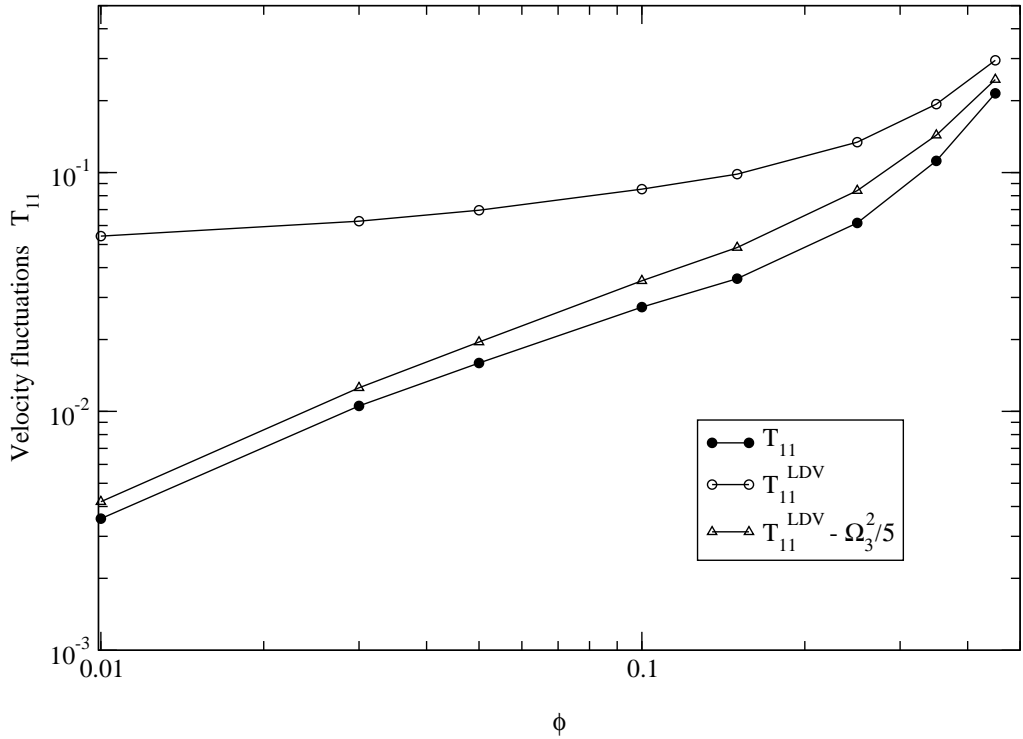}
\caption[]{Particle velocity fluctuations in the mean flow direction, $T_{11}$, as computed directly from the
numerical simulations (solid circles), as if they were measured using the LDV technique (open circles) and 
after correcting the LDV measurements to account for the average rotation in the vorticity direction
(open triangles).}
\label{Correcting_LDV}
\end{figure}

It is known that, due to the spatial but random distribution of the scattering sites within the spheres, 
LDV measurements contain spurious contributions to the linear particle velocity, resulting from the rotation 
of the particles, which are invariably neglected. This would appear to be permissible for the case of the 
temperature measurements, given that most of these spurious contributions average to zero and do not affect 
the variance of the velocity fluctuations because the location of the scattering sites is uncorrelated from
one particle to another. In addition, \cite{LyonL98i} estimated the contribution from the mean particle rotation 
to be one order of magnitude lower than the velocity fluctuations resulting from interparticle interactions, 
and based upon this argument neglected its effect. \cite{ShapleyAB02} explicitly computed the contribution of 
the average rotation of the spheres to the measured velocity fluctuations, but also concluded that its magnitude 
was negligible compared to the fluctuations resulting from collisions between particles. However, the spurious 
contributions to the measured velocity fluctuations originating from the mean angular velocity in the vorticity 
direction are independent of concentration in the dilute limit, and moreover, we have shown earlier that, again 
in the dilute limit, the angular velocity fluctuations are proportional to the volume fraction. Thus, it is clear 
that the spurious contribution to the measured velocity fluctuations due to the angular rotation of the spheres 
eventually becomes important, and even dominant, at low enough concentrations. 
For example, if the scattering sites were distributed uniformly inside the spheres, which rotate with mean angular
velocity $\langle \omega_3 \rangle$, the measured SD of the velocity fluctuations in the direction of the flow, 
$T_{11}^{LDV}$, can be written as, 
\begin{equation}
T_{11}^{LDV} = T_{11} + \frac{1}{5} \left( \Omega_{22} + \Omega_{33} + \langle \omega_3 \rangle^2 \right),
\end{equation}
where the second term on the right hand side corresponds to the above mentioned spurious contribution to the
measured velocity fluctuations due to the rotation of the particles, and is usually neglected.
But since, in the dilute limit, 
$T_{11} + ({1}/{5}) ( \Omega_{22} + \Omega_{33}) = [t_{11}+(1/5)(\tilde\Omega_{22}+\tilde\Omega_{33})] \phi$
(c.f. Eq. \ref{Wvariance}), while $\langle \omega_3 \rangle = \Omega_3 = -1/2$,
it is clear that, at low concentrations, the mean angular rotation of the spheres dominates the
contribution to $T_{11}^{LDV}$, hence, subtracting $\Omega_3^2/5=1/20$ is a correction needed in this limit. 
To illustrate this result we compare, in figure \ref{Correcting_LDV}, 
the velocity fluctuations in the direction of the flow as would have been obtained from the LDV measurements, 
$T_{11}^{LDV}$, and that from the same measurements but corrected by the average rotation, 
$T_{11}^{LDV}-\Omega_3^2/5$, with the real temperature $T_{11}$.
Surprisingly, this simple correction to the analysis of the data remains important even at
concentrations as large as 20\%, or even larger, and in fact the corrected values for $T_{11}^{LDV}$
stay very close to the true temperature $T_{11}$ over the whole range of concentrations investigated.

\section{Summary}

The velocity fluctuations that occur when a simple shear is imposed in a macroscopically homogeneous suspension 
of neutrally buoyant, non-Brownian spheres, and their dependence on the microstructure developed by the
suspensions, were investigated in the limit of vanishingly small Reynolds numbers by means of Stokesian 
dynamics simulations. 
These simulations account for the hydrodynamic interactions between spheres, and also include
a short-range repulsion force that qualitatively models the effects of surface roughness and Brownian forces.
We simulated the evolution of a large number of independent initial hard-sphere random distributions for
strains $\dot\gamma t \sim 100$, which in our previous work, proved sufficiently long to allow us to study 
the system in the asymptotic, fully developed steady state \cite[][]{acrivos1}.

We first discussed the angular structure developed by the suspension undergoing simple shear, and 
showed that, even for exceedingly short ranged interparticle forces, the distribution of particles 
is fore-aft asymmetric at large concentrations, with a depletion 
of pairs oriented in the receding side of the reference particle.
On the other hand, we showed that the distribution of close particles recovered its expected fore-aft symmetry
at low concentrations, but that it still remained anisotropic, with a depletion of pairs oriented close
to the flow direction. 
We were able to accurately describe the observed anisotropy in the pair distribution
function by supposing that permanent doublets were completely absent.
We then showed that the pair distribution function obtained by \cite{batchelor72b} in the dilute limit,
$g_{\text{\tiny BG}}(r)$, accurately follows the simulation results over a wide range of $r$, 
including the large increase in the 
probability of finding pairs of spheres near contact, corresponding to $r \sim 2$. 
However, $g_{\text{\tiny BG}}(r)$ does not take into account the depletion of permanent doublets, 
and it is therefore unable to capture the behavior of the distribution in the limit $r \to 2$.
In fact, in contrast to the divergent behavior of $g_{\text{\tiny BG}}(r)$ for $r \to 2$, the numerical
results suggest that $g(r) \sim 0$ for $r$ less than the minimum distance of approach between two spheres 
in the region of open trajectories ($r_{min} \sim 4\times10^{-5}$).

For the velocity fluctuations, we showed that, for an isotropic configurational probability 
of particles surrounding a reference sphere located at $\boldsymbol r$, 
$P(\mathcal C_N|\boldsymbol r)$, the temperature tensor
is diagonal and that the temperatures in the plane of shear are equal. 
Moreover, we showed that in the dilute limit, the temperature components are proportional to the
volume fraction, and that the temperature in the plane of shear is larger than that
in the vorticity direction. 
Then, by averaging the velocity fluctuations originated in the hydrodynamic interactions between two
spheres, weighted by $g_{\text{\tiny BG}}(r)$, and neglecting to the first approximation the effects 
of the permanent doublets, we computed the temperature tensor in the dilute limit, and found good 
agreement with the results of our numerical simulations, even for moderately concentrated suspensions.
Furthermore, we were able to accurately reproduce the whole velocity autocorrelation function in
both transverse directions, on the basis of only two-particle hydrodynamic interactions.
In contrast, larger discrepancies were found between the corresponding results for the angular velocity 
fluctuations and those obtained in the numerical simulations.
However, in this case we provided a rough estimate for the lower bound of the fluctuations in the
dilute limit.

In order to further investigate the effects of the microstructure on the temperature tensor, we performed
numerical computations in which we calculated the velocity fluctuations for a hard-sphere distribution of
particles subject to the same simple shear flow.
We also calculated the asymptotic behavior in the dilute limit, using a uniform pair 
distribution function $g_{\text{\tiny HS}}(r)=1$, and obtained an excellent agreement with the numerical results
for all the linear and angular velocity fluctuations. 

In addition to the temperature tensor, we presented the full probability distribution of the velocity
fluctuations for both the linear and the angular velocities, in all directions and for three different volume
fractions. We observed different functional forms as the concentration decreases, from a Gaussian 
to an Exponential and finally to a Stretched Exponential form.

Finally, we presented a simple correction term, which only depends on the mean angular velocity of the 
spheres in the vorticity direction, which enhances the interpretation of the LDV measurements at intermediate
and low volume fractions. 

\begin{acknowledgements}
We wish to thank Professor J.~F.~Brady for the use of his simulation codes. 
G. D. thanks I. Baryshev for their helpful comments. 
A.A. and G.D. were partially supported by the Engineering Research Program, Office of Basic Energy and Sciences, 
U.S. Department of Energy under Grant DE-FG02-90ER14139;
G. D. was partially supported by CONICET Argentina and The University of Buenos Aires. 
J.K. was supported by the NASA, Office of Physical and Biological Research under Grant NAG3-2335
and by the Geosciences Program, Office of Basic Energy and Sciences, 
U.S. Department of Energy under Grant DE-FG02-93ER14327;
Computational facilities were provided by the National Energy Resources Scientific Computer Center.
\end{acknowledgements}

\clearpage
\newpage

\bibliography{../../LaTeX/article,../../LaTeX/book,../../LaTeX/mios,../../LaTeX/DownloadedFilesDB/downloaded}

\begin{thebibliography}{27}
\expandafter\ifx\csname natexlab\endcsname\relax\def\natexlab#1{#1}\fi

\bibitem[Averbakh {\em et~al.\/}(1997)Averbakh, Shauly, Nir \&
  Semiat]{AverbakhSNS97}
{\sc Averbakh, A., Shauly, A., Nir, A. \& Semiat, R.} 1997 Slow viscous flows
  of highly concentrated suspensions. i. laser-doppler velocimetry in
  rectangular ducts. {\em Int. J. Multiph. Flow\/} {\bf 23}~(3), 409--424.

\bibitem[Batchelor(1972)]{batchelor72}
{\sc Batchelor, G.~K.} 1972 Sedimentation in a dilute dispersion of spheres.
  {\em J. Fluid Mech.\/} {\bf 52}~(2), 245--268.

\bibitem[Batchelor \& Green(1972{\natexlab{{\em a\/}}})]{batchelor72a}
{\sc Batchelor, G.~K. \& Green, J.~T.} 1972{\natexlab{{\em a\/}}} The
  hydrodynamic interaction of two small freely-moving spheres in a linear flow
  field. {\em J. Fluid Mech.\/} {\bf 56}~(2), 375--400.

\bibitem[Batchelor \& Green(1972{\natexlab{{\em b\/}}})]{batchelor72b}
{\sc Batchelor, G.~K. \& Green, J.~T.} 1972{\natexlab{{\em b\/}}} The
  determination of the bulk stress in a suspension of spherical particles to
  order $c^2$. {\em J. Fluid Mech.\/} {\bf 56}~(3), 401--427.

\bibitem[Bishop {\em et~al.\/}(2002)Bishop, Popel, Intaglietta \&
  Johnson]{BishopPIJ02}
{\sc Bishop, J.~J., Popel, A.~S., Intaglietta, M. \& Johnson, P.~C.} 2002
  Effect of aggregation and shear rate on the dispersion of red blood cells
  flowing in venules. {\em Am. J. Physiol.-Heart Circul. Physiol.\/} {\bf 283},
  H1985--H1996.

\bibitem[Bossis \& Brady(1984)]{BossisB84}
{\sc Bossis, G. \& Brady, J.~F.} 1984 Dynamic simulation of sheared
  suspensions. i. general method. {\em J. Chem. Phys.\/} {\bf 80}~(10),
  5141--5154.

\bibitem[Brady(2001)]{Brady01}
{\sc Brady, J.~F.} 2001 Computer simulation of viscous suspensions. {\em Chem.
  Eng. Sci.\/} {\bf 56}, 2921--2926.

\bibitem[Brady \& Bossis(1988)]{brady88}
{\sc Brady, J.~F. \& Bossis, G.} 1988 Stokesian dynamics. {\em Annu. Rev. Fluid
  Mech.\/} {\bf 20}, 111--140.

\bibitem[Cullen {\em et~al.\/}(2000)Cullen, Duffy, O'Donnell \&
  O'Callaghan]{CullenDOO00}
{\sc Cullen, P.~J., Duffy, A.~P., O'Donnell, C.~P. \& O'Callaghan, D.~J.} 2000
  Process viscometry for the food industry. {\em Trends Food Sci. Technol.\/}
  {\bf 11}, 451--457.

\bibitem[da~Cunha \& Hinch(1996)]{daCunhaH96}
{\sc da~Cunha, F.~R. \& Hinch, E.~J.} 1996 Shear-induced dispersion in a dilute
  suspension of rough spheres. {\em J. Fluid Mech.\/} {\bf 309}, 211--223.

\bibitem[Drazer {\em et~al.\/}(2002)Drazer, Koplik, Khusid \&
  Acrivos]{acrivos1}
{\sc Drazer, G., Koplik, J., Khusid, B. \& Acrivos, A.} 2002 Deterministic and
  stochastic behaviour of non-brownian spheres in sheared suspensions. {\em J.
  Fluid Mech.\/} {\bf 460}, 307--335.

\bibitem[Gadala-Maria \& Acrivos(1980)]{Gadala-MariaA80}
{\sc Gadala-Maria, F. \& Acrivos, A.} 1980 Shear-induced structure in a
  concentrated suspension of solid spheres. {\em J. Rheol.\/} {\bf 24}~(6),
  799--814.

\bibitem[Gotz {\em et~al.\/}(2003)Gotz, Zick \& Kreibich]{gotz2003}
{\sc Gotz, J., Zick, K. \& Kreibich, W.} 2003 Possible optimisation of pastes
  and the accroding apparatus in process engineering by MRI flow experiments.
  {\em Chem. Eng. Proc.\/} {\bf 42}, 517--534.

\bibitem[Husband \& Gadala-Maria(1987)]{HusbandG87}
{\sc Husband, D.~M. \& Gadala-Maria, F.} 1987 Anisotropic particle distribution
  in dilute suspensions of solid spheres in cylindrical couette flow. {\em J.
  Rheol.\/} {\bf 31}~(1), 95--110.

\bibitem[van Kampen(1987)]{vankampen}
{\sc van Kampen, N.~G.} 1987 {\em Stochastic Processes in Physics and
  Chemistry\/}. North-Holland.

\bibitem[Kim \& Karrila(1991)]{kim}
{\sc Kim, S. \& Karrila, S.~J.} 1991 {\em Microhydrodynamics: Principles and
  Selected Applications\/}. Butterworth-Heinemann.

\bibitem[Kolli {\em et~al.\/}(2002)Kolli, Pollauf \& Gadala-Maria]{KolliPG02}
{\sc Kolli, V.~G., Pollauf, E.~J. \& Gadala-Maria, F.} 2002 Transient normal
  stress response in a concentrated suspension of spherical particles. {\em J.
  Rheol.\/} {\bf 46}~(1), 321--334.

\bibitem[Leal(1992)]{leal}
{\sc Leal, L.~G.} 1992 {\em Laminar Flow and Convective Transport Processes\/}.
  Butterworth-Heinemann.

\bibitem[Lyon \& Leal(1998)]{LyonL98i}
{\sc Lyon, M.~K. \& Leal, L.~G.} 1998 An experimental study of the motion of
  concentrated suspensions in twodimensional channel flow. part 1. monodisperse
  systems. {\em J. Fluid Mech.\/} {\bf 363}, 25--56.

\bibitem[Marchioro \& Acrivos(2001)]{marchioro2001}
{\sc Marchioro, M. \& Acrivos, A.} 2001 Shear-induced particle diffusivities
  from numerical simulations. {\em J. Fluid Mech.\/} {\bf 443}, 101.

\bibitem[Parsi \& Gadala-Maria(1987)]{ParsiG87}
{\sc Parsi, F. \& Gadala-Maria, F.} 1987 Fore-and-aft asymmetry in a
  concentrated suspension of solid spheres. {\em J. Rheol.\/} {\bf 31}~(8),
  725--732.

\bibitem[Rampall {\em et~al.\/}(1997)Rampall, Smart \& Leighton]{RampallSL97}
{\sc Rampall, I., Smart, J.~R. \& Leighton, D.~T.} 1997 The influence of
  surface roughness on the particle-pair distribution function of dilute
  suspensions of non-colloidal spheres in simple shear flow. {\em J. Fluid
  Mech.\/} {\bf 339}, 1--24.

\bibitem[Shapley {\em et~al.\/}(2002)Shapley, Armstrong \& Brown]{ShapleyAB02}
{\sc Shapley, N.~C., Armstrong, R.~C. \& Brown, R.~A.} 2002 Laser doppler
  velocimetry measurements of particle velocity fluctuations in a concentrated
  suspension. {\em J. Rheol.\/} {\bf 46}, 241--272.

\bibitem[Shauly {\em et~al.\/}(1997)Shauly, Averbakh, Nir \&
  Semiat]{ShaulyANS97}
{\sc Shauly, A., Averbakh, A., Nir, A. \& Semiat, R.} 1997 Slow viscous flows
  of highly concentrated suspensions. ii. particle migration, velocity and
  concentration profiles in rectangular ducts. {\em Int. J. Multiph. Flow\/}
  {\bf 23}~(4), 613--629.

\bibitem[Sreenivasan(1999)]{sreenivasan99}
{\sc Sreenivasan, K.~R.} 1999 Fluid turbulence. {\em Rev. Mod. Phys.\/} {\bf
  71}~(2), S383--S395.

\bibitem[Voltz {\em et~al.\/}(2002)Voltz, Nitschke, Heymann \&
  Rehberg]{VoltzNHR02}
{\sc Voltz, C., Nitschke, M., Heymann, L. \& Rehberg, I.} 2002 Thixotropy in
  macroscopic suspensions of spheres. {\em Phys. Rev. E\/} {\bf 65}, 051402.

\bibitem[Wang {\em et~al.\/}(1996)Wang, Mauri \& Acrivos]{wang96}
{\sc Wang, Y., Mauri, R. \& Acrivos, A.} 1996 The transverse shear-induced
  liquid and particle tracer diffusivities of a dilute suspension of spheres
  undergoing a simple shear flow. {\em J. Fluid Mech.\/} {\bf 327}, 255--272.

\end{thebibliography}
\bibliographystyle{jfm}

\end{document}